\documentclass[fleqn,usenatbib]{mnras}

\usepackage{newtxtext,newtxmath}

\usepackage[T1]{fontenc}
\usepackage{ae,aecompl}

\usepackage{graphicx}	
\usepackage{amsmath}	
\usepackage{bm}

\newcommand{\Mpcph}{\,h^{-1}{\rm Mpc}}
\newcommand{\hpMpc}{\,{\rm Mpc}^{-1}h}

\newcommand{\orcid}[1]{\href{https://orcid.org/#1}{\includegraphics[width=0.7em]{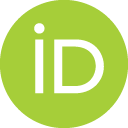}}}

\title[Testing GR with quasar and CMB lensing]{Testing General Relativity on cosmological scales at redshift $z \sim 1.5$ with quasar and CMB lensing}

\author[Y. Zhang et al.]{\parbox{\textwidth}{
Yucheng Zhang \orcid{0000-0002-9300-2632},$^{1}$\thanks{E-mail: \href{mailto:yucheng.zhang@nyu.edu}{yucheng.zhang@nyu.edu}}
Anthony R. Pullen,$^{1,2}$
Shadab Alam \orcid{0000-0002-3757-6359},$^{3}$
Sukhdeep Singh,$^{4}$\\
Etienne Burtin,$^{5}$
Chia-Hsun Chuang,$^{6}$
Jiamin Hou,$^{7}$
Brad W. Lyke,$^{8}$
Adam D. Myers,$^{8}$
Richard Neveux,$^{5}$
Ashley J. Ross,$^{9}$
Graziano Rossi,$^{10}$
Cheng Zhao$^{11}$
}\\\\
$^{1}$ Center for Cosmology and Particle Physics, Department of Physics, New York University, 726 Broadway, New York, NY 10003, USA \\
$^{2}$ Center for Computational Astrophysics, Flatiron Institute, New York, NY 10010, USA \\
$^{3}$ Institute for Astronomy, University of Edinburgh, Royal Observatory, Blackford Hill, Edinburgh, EH9 3HJ, UK \\
$^{4}$ Berkeley Center for Cosmological Physics, University of California, Berkeley, CA 94720, USA \\
$^{5}$ IRFU,CEA, Universit\'e Paris-Saclay, F-91191 Gif-sur-Yvette, France \\
$^{6}$ Kavli Institute for Particle Astrophysics and Cosmology, Stanford University, 452 Lomita Mall, Stanford, CA 94305, USA \\
$^{7}$ Max-Planck-Institut f\"ur Extraterrestrische Physik, Postfach 1312, Giessenbachstrasse 1, 85748 Garching bei M\"unchen, Germany \\
$^{8}$ University of Wyoming, 1000 E. University Ave., Laramie, WY 82071, USA \\
$^{9}$ Center for Cosmology and Astro-Particle Physics, Ohio State University, Columbus, Ohio, USA \\
$^{10}$ Department of Physics and Astronomy, Sejong University, Seoul, 143-747, Korea \\
$^{11}$ Institute of Physics, Laboratory of Astrophysics, \'Ecole Polytechnique F\'ed\'erale de Lausanne (EPFL), Observatoire de Sauverny, 1290 Versoix, Switzerland
}

\date{Accepted XXX. Received YYY; in original form ZZZ}

\pubyear{2020}

\begin{document}
\label{firstpage}
\pagerange{\pageref{firstpage}--\pageref{lastpage}}
\maketitle

\begin{abstract}
We test general relativity (GR) at the effective redshift $\bar{z} \sim 1.5$ by estimating the statistic $E_G$, a probe of gravity, on cosmological scales $19 - 190\Mpcph$. This is the highest redshift and largest scale estimation of $E_G$ so far. We use the quasar sample with redshifts $0.8 < z < 2.2$ from Sloan Digital Sky Survey IV extended Baryon Oscillation Spectroscopic Survey Data Release 16 as the large-scale structure (LSS) tracer, for which the angular power spectrum $C_\ell^{qq}$ and the redshift-space distortion parameter $\beta$ are estimated. By cross-correlating with the $\textit{Planck}$ 2018 cosmic microwave background (CMB) lensing map, we detect the angular cross-power spectrum $C_\ell^{\kappa q}$ signal at $12\,\sigma$ significance. Both jackknife resampling and simulations are used to estimate the covariance matrix (CM) of $E_G$ at five bins covering different scales, with the later preferred for its better constraints on the covariances. We find $E_G$ estimates agree with the GR prediction at $1\,\sigma$ level over all these scales. With the CM estimated with $300$ simulations, we report a best-fitting scale-averaged estimate of $E_G(\bar{z})=0.30\pm 0.05$, which is in line with the GR prediction $E_G^{\rm GR}(\bar{z})=0.33$ with $\textit{Planck}$ 2018 CMB+BAO matter density fraction $\Omega_{\rm m}=0.31$. The statistical errors of $E_G$ with future LSS surveys at similar redshifts will be reduced by an order of magnitude, which makes it possible to constrain modified gravity models.
\end{abstract}

\begin{keywords}
cosmology: theory -- cosmology: observations -- large-scale structure of Universe -- gravitation -- gravitational lensing: weak -- cosmic background radiation
\end{keywords}



\section{Introduction} \label{sec:intro}
The expansion of the universe was first discovered by measuring the redshifts and relative distances of galaxies~\citep{Hubble1929}. One of the milestones in cosmology in the past decades has been the detection of a negative deceleration parameter and hence the accelerated expansion of the Universe at late times from supernovae observations~\citep{Riess1998,Perlmutter1999}. Many theoretical models of cosmology and gravity have been proposed to explain the cosmic expansion and acceleration~\citep[see e.g.][and references therein]{Silvestri2009}, among which $\Lambda$-cold dark matter ($\Lambda$CDM) has been regarded as the standard model for its simplicity and success in explaining a wide range of cosmological observations, including the CMB (cosmic microwave background) surveys~\citep[e.g.][]{Planck2018-cosmo} and galaxy redshift surveys~\citep[e.g.][]{Alam2017-BOSS}. $\Lambda$CDM takes general relativity (GR) as the true theory for gravity on both galactic and cosmological scales, and assumes the existence of the cosmological constant ($\Lambda$), a special form of dark energy (DE) whose spatially uniform energy density does not evolve with cosmic expansion, and CDM, along with ordinary (baryonic) matter. Although the expansion history can be well described by $\Lambda$CDM-GR by fine-tuning the relative density ratios of the energy components, the nature of dark matter (DM) and DE are not well understood and their properties are hard to detect with observations. On the other hand, some modified gravity (MG) models~\citep[see e.g.][]{Carroll2005, Sotirious2010, Dvali2000}, which can predict the same expansion history of the universe as $\Lambda$CDM-GR with or completely without assuming the existence of DE, have been developed to challenge GR as the true theory for gravity on cosmological scales. There have been some great reviews of the two approaches, see e.g. \citet{Peebles2003} for the cosmological constant and DE, \citet{Clifton2012} for MG, and \citet{Joyce2016} for a comparison.

Despite the degeneracy between $\Lambda$CDM-GR and MG models in explaining the cosmic expansion, their predictions of the growth of the DM large scale structure (LSS) are usually distinguishable. Combining the gravitational lensing $\nabla^2(\Psi - \Phi)$ and the divergence of the peculiar velocity $\theta$, \citet{Zhang2007} proposed a statistic $E_G$ as a function of redshift and scale, to probe gravity on cosmological scales. Lensing is related to the underlying matter overdensity $\delta$ through the Poisson equation which depends on the gravity model~\citep[see e.g.][]{Hojjati2011}. On linear scales, $\theta=-f\delta$, where $f$ is the linear growth rate. In real surveys, instead of the DM field, the direct observables are the LSS tracers, e.g. galaxies or quasars. The distribution of these tracers is connected to the underlying matter perturbation field with the clustering bias $b$, which varies with the physical properties of the tracers that are targeted in a particular survey. Defined as the ratio between $\nabla^2(\Psi - \Phi)$ and $\theta$, $E_G$ has the advantage of being independent of $b$ and the variance of the matter density field $\sigma_8$. Interested readers may refer to~\citet{Ishak2019} for a comprehensive review on various cosmological tests of GR, including the $E_G$ statistic.

The estimation of $E_G$ requires data from both gravitational lensing and redshift surveys. Accurate estimates of tracers' redshifts are necessary in order to do the 3-D clustering analysis, from which the growth of the structure can be probed. Thus spectroscopic redshift surveys are usually preferred. For photometric surveys, \citet{Giannantonio2016} proposed a statistic $D_G$, which does not require the estimation of the growth rate. However, this quantity cannot be directly used to discriminate GR and MG models. Using galaxy-galaxy lensing and luminous red galaxies (LRGs), $E_G$ has been measured over scales $\lesssim 70\Mpcph$ at redshifts in $0.2< z <0.6$~\citep{Reyes2010, Blake2015, delaTorre2017, Alam2017, Amon2018, Singh2018,Blake2020}. Besides tracing the lensing signal with background galaxies, \citet{Pullen2015} proposed to use the cosmic microwave background (CMB) lensing map, which allows the estimation of $E_G$ at higher redshifts and larger scales~\citep{Pullen2016, Singh2018}.

In this work, using quasars and CMB lensing, we test $\Lambda$CDM-GR on cosmological scales $19-190\Mpcph$ at the effective redshift $\bar{z} \sim 1.5$, which is the highest redshift and largest scale $E_G$ estimation so far. Quasars, also known as quasi-stellar objects (QSOs), are active galactic nuclei (AGN) with very high luminosity, which makes them good candidates to trace LSS at higher redshifts (e.g. $1<z<2$). As part of the primary motivation of constraining $E_G$, we also investigate the reliability of quasars as a tracer of the DM in both auto- and cross-clustering analyses. The redshift range of the quasar targets is very close to the peak of CMB lensing kernel at $z\sim 2$, so we should expect a promising cross-correlation signal, which is usually harder to be detected than the auto-correlation.
Assumptions of the cosmology and gravity models have to be made in order to do certain estimations and generate the simulations needed. So for now it is very difficult to design one blind test for various gravity models. To do a rigorous estimation of $E_G$ based on other MG models, the corresponding changes have to be made for either simulations or analytic calculations~\citep[see e.g.][]{Hojjati2011}.

The paper is organized as follows. In Section~\ref{sec:eg}, we review the $E_G$ theory and describe the estimator we use. The quasar and CMB data, simulations and jackknife resampling for the estimation of covariance matrices are described in Section~\ref{sec:data}. Section~\ref{sec:cls} includes analytic models, estimators, systematics and calibrations for the angular power spectra. Section~\ref{sec:rsd} describes our estimation of the quasar 2-point correlation function and the maximum likelihood fitting of the redshift-space distortion (RSD) parameter. We present all the estimates and our final results in Section~\ref{sec:results} and conclude in Section~\ref{sec:conclusions}.

For our self-consistency test of GR, wherever needed, we assume a flat $\Lambda$CDM fiducial cosmology with \textit{Planck} 2018 CMB$+$BAO parameters~\citep{Planck2018-cosmo}: $\Omega_{\rm m}=0.3111\pm 0.0056$, $\Omega_{\rm c} h^2=0.11933\pm 0.00091$, $\Omega_{\rm b} h^2=0.02242\pm 0.00014$, $n_{\rm s}=0.9665\pm 0.0038$, $H_0=67.66\pm 0.42$, and $\sigma_8=0.8102\pm 0.0060$.

\section{\texorpdfstring{$E_G$}{EG} formalism and estimator} \label{sec:eg}
In this section, we briefly review the $E_G$ theory and describe the estimator used in this work. We assume a flat Universe described by the perturbed Friedmann-Robertson-Walker (FRW) metric in conformal Newtonian gauge,
\begin{equation}
    ds^2 = a(\tau)\left[(1+2\Psi)d\tau^2 - (1+2\Phi)dx^2\right]\, ,
\end{equation}
where $\Psi$ and $\Phi$ are the scalar perturbations to the time and spatial components of the metric. The statistic $E_G$ is defined in Fourier Space~\citep{Zhang2007} as
\begin{equation}
    \begin{split}
    E_G(k, z) &= \left[\frac{\nabla^2(\Psi - \Phi)}{-3H_0^2(1+z)\theta}\right]_k \\
    &= \frac{k^2(\Psi - \Phi)}{3H_0^2(1+z)\theta}\, ,
    \end{split}
\end{equation}
where $H_0$ is the Hubble constant and $\theta = \nabla\cdot\bm{v} / H(z)$ is the divergence of the comoving peculiar velocity field. In linear perturbation theory, $\theta = - f\delta$, where $f$ is the linear growth rate and $\delta$ is the matter perturbation. For GR, assuming no anisotropic stress ($\Phi = -\Psi$) and using Poisson equation $\nabla^2\Psi = 4\upi Ga^2\rho_{m}\delta$, we have
\begin{equation}
    E^{\rm GR}_G(z) = \frac{\Omega_{m,0}}{f(z)}\, ,
\label{eq:eg_gr}
\end{equation}
where $\Omega_{m,0} = \rho_{m,0} / \rho_{\rm crit,0}$ is the fraction of matter density today with $\rho_{\rm crit,0} = 3H_0^2 / 8\upi G$, and $f(z)\simeq \Omega_m(z)^{\gamma}$ with $\gamma\simeq 0.55$ and
\begin{equation}
    \Omega_m(z) = \frac{\Omega_{m,0} (1+z)^3}{\Omega_{m,0} (1+z)^3 + (1-\Omega_{m,0})}
\end{equation}
at late times. Notice that $E^{\rm GR}_G(z)$ is scale-independent and only relies on the relative fraction of matter density in the Universe. For different MG models, $E_G$ can have different amplitudes or be scale-dependent~\citep{Zhang2007,Pullen2015}.

The angular estimator for $E_G$ at the effective redshift $\bar{z}$ can be constructed as~\citep{Pullen2015}
\begin{equation}
    \begin{split}
    \left.\hat{E}_G(\ell)\right|_{\bar{z}} &= \left.\frac{c^2}{3H_0^2}\frac{C_\ell^{\kappa q}}{C_\ell^{\theta q}}\right|_{\bar{z}} \\
    &\simeq \Gamma(\bar{z})\frac{C_\ell^{\kappa q}}{\beta(\bar{z})C_\ell^{qq}} \, ,
    \end{split}
\label{eq:eg_est}
\end{equation}
where $c$ is the speed of light, $\kappa$ and $q$ denote the CMB lensing convergence and quasar overdensity maps respectively. $C_\ell$'s are the angular power spectra, $\beta=f/b$ is the RSD parameter given by the ratio of the linear growth rate and the clustering bias, and $\Gamma$ is an analytic factor,
\begin{equation}
    \Gamma(\bar{z}) = \frac{2c}{3H^2_0}\frac{H(\bar{z})f_q(\bar{z})}{(1+\bar{z})W(\bar{z})}\, ,
\end{equation}
where $f_q(\bar{z})$ is the normalized redshift distribution of the quasar sample at the effective redshift and $W(z)$ is the CMB lensing kernel. $f_q(z)$ and $W(z)$ work as the radial projection kernels for $q$ and $\kappa$ fields when we transform the 3-D power spectra $P(k, z)$ into angular $C_\ell$'s, as shown in Eq.~\ref{eq:clkq} and Eq.~\ref{eq:clqq}. To convert $C_\ell^{\theta q}$ to the directly measurable $C_\ell^{qq}$, the approximation made in Eq.~\ref{eq:eg_est} which includes the substitution of a certain redshift-dependent factor with the effective value at $\bar{z}$ is not perfect. This can cause a systematic bias around $5\%$ to our $E_G$ estimation. Following~\citet{Pullen2016} and assuming a scale-independent linear bias $b(z)$, we introduce the calibration factor
\begin{equation}
    C_{\Gamma} = \frac{c}{2}\frac{W(\bar{z})(1+\bar{z})}{H(\bar{z})f_q(\bar{z})}\frac{C_\ell^{mq}}{Q_\ell^{mq}}\, ,
\label{eq:gamma_cor}
\end{equation}
where
\begin{equation}
    C_\ell^{mq} \equiv \int_{z_1}^{z_2}dz \chi^{-2}(z) \frac{H(z)}{c} f^2_q(z) b(z) P_{m}\left(\frac{\ell+1/2}{\chi(z)},z\right)\, ,
\end{equation}
and
\begin{equation}
    Q_\ell^{mq} \equiv \frac{1}{2} \int_{z_1}^{z_2}dz (1+z)\chi^{-2}(z) W(z)f_q(z) b(z) P_{m}\left(\frac{\ell+1/2}{\chi(z)}, z\right)\, ,
\end{equation}
where $\chi(z)$ is the radial comoving distance at redshift $z$, $P_m$ is the matter power spectrum and Limber approximation $k\chi\simeq \ell+1/2$ has been used. Due to the limited size of the quasar sample, it is hard to study the redshift evolution of the bias by cutting the redshift range into a few smaller bins. Here we just take a constant bias at the effective redshift, i.e. $b(z)\simeq b(\bar{z})$. We also tried an eBOSS quasar bias model presented in~\citet{Laurent2017}, and the difference is negligible considering that the systematic bias calibrated by $C_\Gamma$ is only around $5\%$ of the $E_G$ signal. Another systematic bias concern is the non-linear quasar bias and the imperfect connection between quasars and the matter field at small scales, which is hard to model and needs to be corrected with N-body simulations. However, for the scales ($\geq 19\Mpcph$) we are considering, this systematic bias should be negligible~\citep{Pullen2016,Singh2018}.

The correspondence between multipoles $\ell$ and linear scales $\chi_\perp$ at a certain redshift is given by $\chi_\perp = 2\upi\chi(z)/\ell$. With Eq.~\ref{eq:eg_est}, we can estimate $\hat{E}_G(\ell)$ for a range of multipoles. These multipoles are binned into a few bandpowers in practice, with more details discussed in the estimation of $C_\ell$'s (Section~\ref{subsec:cls_est}). In the end, we find the best-fit (denoted as $\bar{E}_G$) of the overall amplitude of $\hat{E}_G(\ell)$ to compare to the GR prediction. To make the discussion coherent, we present our fitting method along with our estimates of the covariance matrix for $E_G(\ell)$ in Section~\ref{subsec:eg_res}.

\section{Data and Covariances} \label{sec:data}
In this section, we describe the quasar and CMB lensing data used in this work. We also discuss the simulations and the jackknife resampling method used to estimate the covariance matrices.

\subsection{Quasar catalogs} \label{subsec:quasar_catalogs}

\begin{figure}
    \centering
    \includegraphics[width=\columnwidth]{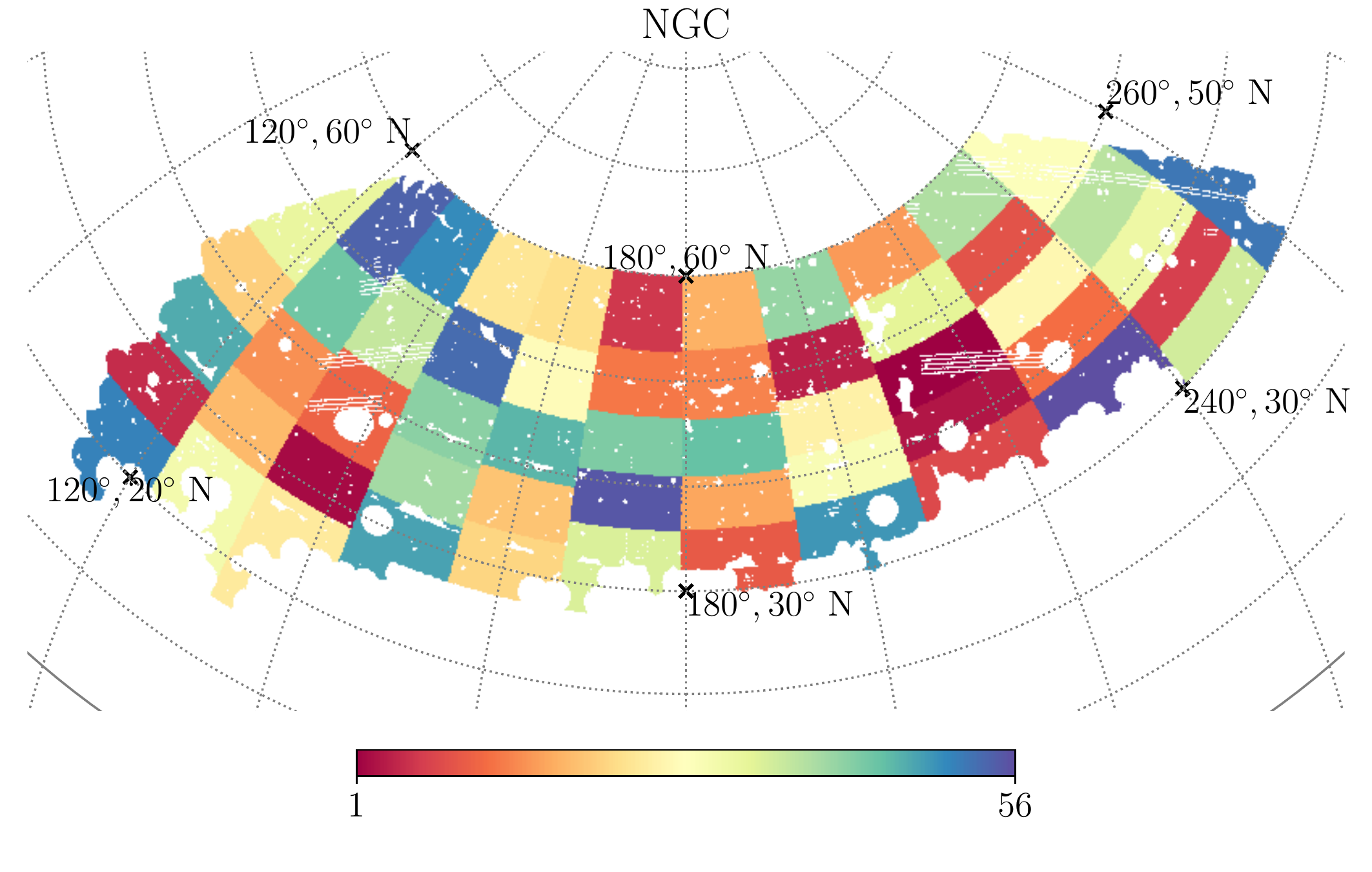} \\
    \includegraphics[width=\columnwidth]{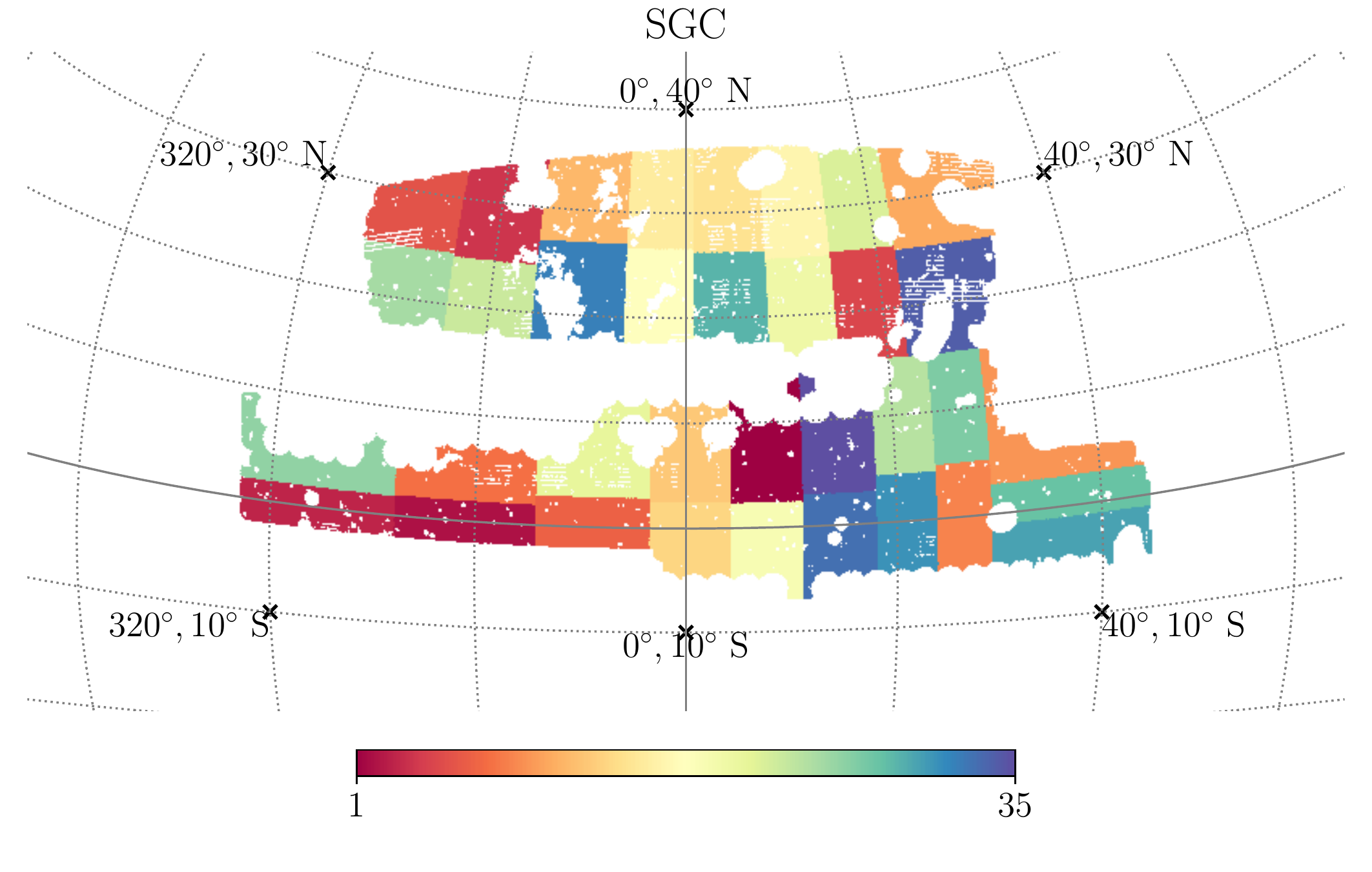}
    \caption{The overlapped sky coverage of \textit{Planck} 2018 CMB lensing and eBOSS DR16 quasar NGC (\textit{upper}) and SGC (\textit{lower}) clustering catalogs. NGC (SGC) covers about $2929$ ($1815$) $\deg^2$. The orientation of the regions are shown in J2000 coordinates. For jackknife resampling, NGC and SGC are divided into $56$ and $35$ equally-weighted regions respectively.}
    \label{fig:mask}
\end{figure}

We use the quasar sample for clustering analysis from the fourth phase of the Sloan Digital Sky Survey (SDSS-IV)~\citep{Blanton2017} extended Baryon Oscillation Spectroscopic Survey (eBOSS)~\citep{Dawson2016} Data Release 16 (DR16)~\citep{SDSS-DR16}, which is observed with the Sloan Foundation 2.5-meter Telescope located at the Apache Point Observatory~\citep{Gunn2006} with double-armed spectrographs~\citep{Smee2013}. The construction of these eBOSS DR16 clustering catalogs for quasars from the complete SDSS DR16 quasar (DR16Q) catalog~\citep{Lyke2020} is described in~\citet{Ross2020}, along with the catalogs for luminous red galaxies (LRGs) and emission line galaxies (ELGs). The quasar sample comprises the north galactic cap (NGC) and the south galactic cap (SGC), which correspond to two separate regions on the sky. Since jackknife resampling is used for covariance estimation (see Section~\ref{subsec:cov}), we only use the sky region covered by both the quasar and CMB lensing surveys (Fig.~\ref{fig:mask}).
The sky coverage fraction and number of quasars are shown in Table~\ref{tab:data}.
\begin{table}
    \centering
    \caption{The overlapped sky coverage fraction of eBOSS DR16 quasar catalogs and \textit{Planck} 2018 CMB lensing, and the corresponding number of quasars. The (weighted) mean and median redshifts agree with each other (see text), denoted as $\bar{z}$. The last column shows the number of quasars which are in the original eBOSS DR16 catalogs but not covered by the \textit{Planck} CMB lensing footprint, and hence these targets are not included in the data analysis.}
    \begin{tabular}{ccccc}
        \hline
        Cap & $f_{\text{sky}}$ ($\%$) & \# quasars & $\bar{z}$ & \# masked\\
        \hline
        NGC & $7.1$ & $210\,881$ & $1.51$ & $7\,328$ \\
        SGC & $4.4$ & $116\,249$ & $1.52$ & $9\,250$ \\
        \hline
    \end{tabular}
    \label{tab:data}
\end{table}
This overlapped coverage masks out around $3.4\,\%$ quasars in NGC and $7.4\,\%$ quasars in SGC. Even without jackknife resampling, using this total mask is still reasonable since the removed quasars do not have the corresponding lensing signal anyway.

Using the \texttt{HEALPix}~\citep{Gorski2005} pixelization, we construct the quasar overdensity map with
\begin{equation}
    \delta_i = \frac{n_i}{\bar{n}} - 1\, ,
\end{equation}
where $i$ is the pixel index, $n_i = \sum_{q\in i} w_q$ is the weighted number count of quasars for each pixel and $\bar{n}$ is the the average over all covered pixels. The weight for each quasar is given by $w_q = w_{\rm sys}\cdot w_{\rm cp}\cdot w_{\rm noz}$, where $w_{\rm cp}\cdot w_{\rm noz}$ corrects for the spectroscopic completeness due to close pairs and redshift failures across fibers, and $w_{\rm sys}$ accounts for the imaging systematics. Additionally, for the estimation of the correlation function, $w_{\rm FKP}$ is also applied to optimize the clustering statistics~\citep{Feldman1994}. The determination of all these weights is described in detail in~\citet{Ross2020}.

\begin{figure}
    \centering
    \includegraphics[width=\columnwidth]{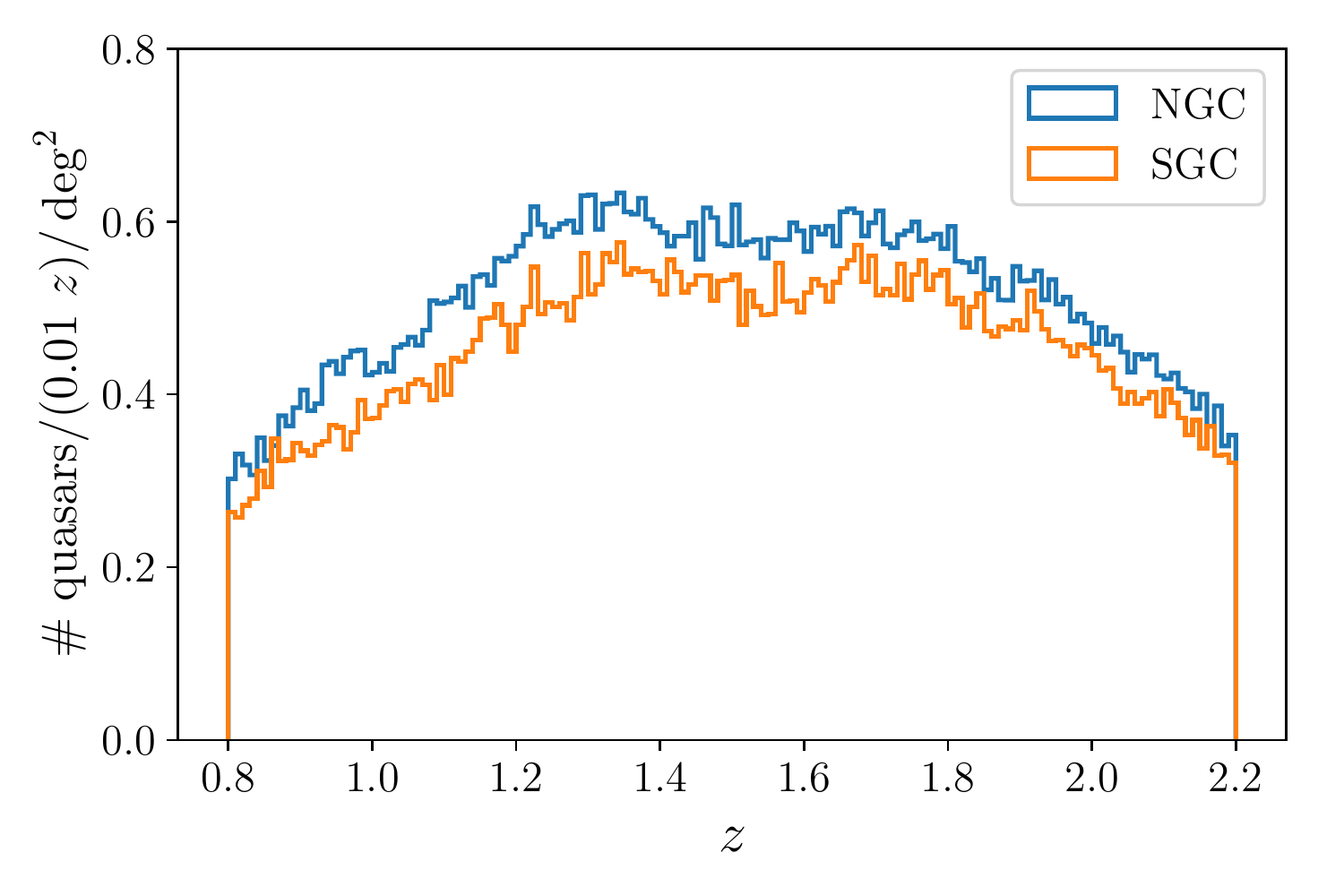}
    \caption{Number density redshift distribution of eBOSS DR16 quasar clustering catalogs (with the overlapped sky coverage with CMB lensing applied, see text). NGC has a higher number density than SGC, which results in lower shot noise.}
    \label{fig:zdist}
\end{figure}

The redshift distribution of the two catalogs are shown in Fig.~\ref{fig:zdist}, where we see that NGC has a higher number density than SGC. So the shot noise due to the Poisson distribution of the quasars, which is inversely proportional to the number density, is lower for NGC than SGC.
The quasars are observed in redshift bin $0.8 < z < 2.2$, for which we need to determine the effective redshift for our angular analysis. The recommended definition of the effective redshift in eBOSS DR16 clustering analysis is given by
\begin{equation}
    z_{\rm eff} = \frac{\sum_{i,j} w_i w_j (z_i + z_j)/2}{\sum_{i,j} w_i w_j}\, ,
    \label{eq:zeff}
\end{equation}
which is proposed for the measurement of the 2-point correlation function and the summation is conducted over pairs with separation distance $25\leq s \leq 120\hpMpc$. With this definition, \citet{Hou2020} find $z_{\rm eff}\simeq 1.48$ for the full clustering quasar sample. For both NGC and SGC quasar samples used in this work, we find that the mean, weighted mean ($\sum_{i}w_i z_i/\sum_i w_i$) and median redshifts agree with each other, with the value shown as $\bar{z}$ in Table~\ref{tab:data}. Although the overlapped mask removes some quasars, these redshift values almost remain the same. The tiny difference in the definitions of the effective redshift is completely negligible compared with the statistical accuracy. Thus for simplicity, in this work, we take the effective redshift at $\bar{z}=1.5$ for both NGC and SGC.

\subsection{CMB lensing map}

The gravitational lensing convergence ($\kappa$) map used is the minimum-variance estimate with CMB temperature and polarization measurements~\citep{Planck2018-lensing}, reconstructed and provided as part of the \textit{Planck} 2018 data release~\citep{Planck2018}. The map covers about $70$ percent of the sky and is provided in spherical harmonics $\kappa_{\ell m}$'s up to $\ell=4096$. However, in this work, we only use the multipoles in $8\leq \ell \leq 2048$. We do not use the multipoles $\ell > 2048$ due to the significant reconstruction noise at those very small scales. Since we are only considering well-defined linear scales $100\leq \ell \leq 1000$ for the angular cross-power spectra estimation of quasar and CMB lensing, contributions from those much smaller and non-linear scales should be negligible compared with the statistical errors.

\subsection{Covariance matrices} \label{subsec:cov}

Covariance matrices (CMs) are needed for constructing the likelihood functions used in the posterior distribution sampling of the parameters, e.g. RSD parameters and scale-averaged $\bar{E}_G$. Like any other statistics, an accurate estimation of the CM relies on a large number of samples. In this work, we estimate the CMs in two ways. One is using simulations, and the other is jackknife resampling, which only depends on the data itself.

For simulations, we run all of them through the same data analysis pipeline as we do for the real data, with which we can then construct the CM for any statistical quantity in the procedure. We use $300$ simulated $\kappa$ maps coming with \textit{Planck} 2018 CMB lensing analysis~\citep{Planck2018-lensing}, in which the lensing reconstruction noise is included. For the eBOSS quasar sample, \citet{Zhao2020} generated $1000$ effective Zel'dovich approximation mock catalogs~\citep[EZ mocks,][]{Chuang2015}. The fiducial cosmology for generating the mocks is flat $\Lambda$CDM with parameters: $\Omega_{\rm m} = 0.307115$, $\Omega_{\rm b} = 0.048206$, $h = 0.6777$, $\sigma_8 = 0.8225$, and $n_{\rm s} = 0.9611$. These are slightly different from the \textit{Planck} 2018 CMB+BAO parameters we are assuming, but the influence on the CMs should be negligible. Combining these simulated $\kappa$ maps and EZ mocks, we have $300$ sets of independent simulations for our $E_G$ analysis. These lensing maps and quasar mocks are not correlated, which results in zero mean signal and lower error estimates (Eq.~\ref{eq:clkq_err}) for the cross correlation $C_\ell^{\kappa q}$. As discussed in Section~\ref{subsec:cls_res}, the contribution of $C_\ell^{\kappa q}$ signal to the error distribution of itself is negligible compared with the noise level in the auto correlation of the current surveys. However, the $C_\ell^{\kappa q}$ signal is important in the CM estimation for functions of it like $E_G$, which can be seen from the Gaussian error propagation. We discuss our approach to fix this issue in Section~\ref{subsec:eg_res}, where the estimates of CMs for $E_G(\ell)$ are presented. Although using \textit{realistic} simulations is a reliable way to estimate CMs since we can run as many simulations as needed (with enough computing resources), it should still be reiterated that simulations depend on the fiducial model, where extra consideration is necessary for the purpose of testing different models on the data.

Another CM estimation method which only relies on the data sample is jackknife resampling. In this work, we divide the overlapped sky coverage of quasar and CMB lensing into $N$ equally weighted regions and make \textit{leave-one-out} jackknife samples by taking one region out each time. This process leaves us $N$ correlated re-samples of the original full data. We do the analysis for each of these jackknife samples, with each result denoted as a vector $\bm{x}$, e.g. the correlation function or power spectrum. Then the covariance matrix of $\bm{x}$ is given by
\begin{equation}
    \mathrm{Cov}(x_i,x_j) = \frac{N-1}{N}\sum_{k=1}^{N}\left(x_i^{(k)}-\bar{x}_i\right)\left(x_j^{(k)}-\bar{x}_j\right)\, ,
\end{equation}
where $\bar{\bm{x}}$ is the mean of all the jackknife estimates, which are labeled with index $k$. Compared with the normal unbiased sample CM estimation, a factor of $(N-1)^2/N$ is multiplied, which corresponds to the fact that the jackknife samples are not independent. Jackknife resampling has the advantage of being dependent only on the data, which hence naturally includes all the systematics and noise in the observations. However, the maximum number of jackknife samples is limited by the largest scale to be probed. In this work, by requiring the linear scale of each region to be at least two times the largest scale we are interested in, we are able to use $56$ ($35$) jackknives for NGC (SGC) (Fig.~\ref{fig:mask}). We make sure that jackknife resampling is unbiased by comparing the mean of all the jackknife estimates with the estimate using the full data sample. It turns out that for the statistics in this work, they are always consistent. However, the number of jackknives used may not be enough to give us accurate estimates of the CMs, especially the off-diagonal terms (i.e. cross correlations between different scales), whose relative strength compared with variances (diagonal terms) can be quantified with the correlation matrix,
\begin{equation}
    \text{Corr}\left(\mathbf{C}\right)_{ij} = \frac{\mathbf{C}_{ij}}{\sqrt{\mathbf{C}_{ii}\mathbf{C}_{jj}}}\,,
\label{eq:cor}
\end{equation}
where $\mathbf{C}$ denotes the CM.

The estimated $\hat{\mathbf{C}}$ for a multivariate Gaussian vector with a limited number of samples follows the Wishart distribution, which is an unbiased estimate of the true CM, $\mathbf{C}$. However, $\hat{\mathbf{C}}^{-1}$, the inverse of $\hat{\mathbf{C}}$, which obeys the inverse Wishart distribution, is a biased estimate of $\mathbf{C}^{-1}$ due to the error in $\hat{\mathbf{C}}$. This can be corrected with a simple factor~\citep{Hartlap2007},
\begin{equation}
    \hat{\mathbf{C}}^{-1}_{\text{unbiased}} = \left(1-\frac{N_d+1}{N_s-1}\right)\hat{\mathbf{C}}^{-1}\, ,
\label{eq:cinv}
\end{equation}
where $N_d$ is the size of the data vector and $N_s$ is the number of samples. Furthermore, the error in $\hat{\mathbf{C}}$ propagates to the CM of the model parameters in the maximum likelihood fitting~\citep{Dodelson2013}. This can be corrected by multiplying the factor 
\begin{equation}
    M = \frac{1+B(N_d-N_p)}{1+A+B(N_p+1)}
\label{eq:mab}
\end{equation}
to the CM of the parameters~\citep{Percival2014}, where $N_p$ is the number of parameters and
\begin{equation}
    \begin{split}
        A &= \frac{2}{(N_s-N_d-1)(N_s-N_d-4)}\, ,\\
        B &= \frac{N_s-N_d-2}{(N_s-N_d-1)(N_s-N_d-4)}\, .
    \end{split}
\end{equation}
It should be noticed that the above corrections are derived for independent samples like the simulations, which may not be the proper solution for jackknife samples~\citep{Taylor2013}. However, more detailed discussion is out of the scope of this paper, which we leave for future work. Specifically in this work, two CMs are used for fitting purposes. One is for the 2-point correlation function of quasars in RSD fitting and the other is for the fitting of $E_G(\ell)$ over scales.

\section{Angular power spectra} \label{sec:cls}
In this section, we describe the theoretical models and estimators for the angular power spectra. We also discuss the influence of possible systematics and the corresponding calibrations applied.

\subsection{Theory} \label{sec:cls-theory}
The analytic expressions for the angular power spectra can be derived by integrating the 3-D power spectra $P(k,z)$ over the wave number $k$, with proper radial projection kernels $F(\chi)$ applied. At high $\ell$'s (e.g. $\ell>10$ is good enough for the wide redshift bin of the quasar sample), the spherical Bessel functions $j_\ell(k\chi)$ vary fast compared with $F(\chi)$, which picks out the scale $k\simeq (\ell+1/2)/\chi(z)$. Based on this, the Limber approximation replaces the $j_\ell$ with the Dirac delta function, which significantly speed up the numerical evaluation of the integral. In what follows, this approximation is always applied.

The CMB lensing $\times$ quasar cross-power spectrum reads
\begin{equation}
    C_\ell^{\kappa q} = \frac{1}{2} \int_{z_1}^{z_2}dz \chi^{-2}(z) W(z) f_q(z) P_{\nabla^2(\Psi-\Phi)q}\left(\frac{\ell+1/2}{\chi(z)},z\right)\, ,
\label{eq:clkq}
\end{equation}
where $\chi(z)$ is the radial comoving distance at redshift $z$, $W(z) = \chi(z)\left[1-\chi(z)/\chi\left(z_{\rm CMB}\right)\right]$ is the CMB lensing kernel with $z_{\rm CMB}\simeq 1100$, $f_q(z) = \frac{1}{N}\frac{dN}{dz}$ is the normalized quasar redshift distribution and $P_{\nabla^2(\Psi-\Phi)q}(k,z)$ is the 3-D cross-power spectrum of the two fields. Assuming GR and using the Poisson equation to replace the lensing convergence with matter perturbation, Eq.~\ref{eq:clkq} can be written as
\begin{equation}
\begin{split}
    C_\ell^{\kappa q} =\ &\frac{3\Omega_{m,0}H_0^2}{2c^2} \int_{z_1}^{z_2}dz \chi^{-2}(z) (1+z) W(z) f_q(z) \\
    &\times P_{mq}\left(\frac{\ell+1/2}{\chi(z)},z\right)\, .
\end{split}
\label{eq:clkq_gr}
\end{equation}
Similarly, the quasar auto-power spectrum is given by
\begin{equation}
    C_\ell^{qq} = \int_{z_1}^{z_2}dz \chi^{-2}(z) \frac{H(z)}{c} f^2_q(z) P_{qq}\left(\frac{\ell+1/2}{\chi(z)},z\right)\, ,
\label{eq:clqq}
\end{equation}
where $H(z)$ is the Hubble parameter at redshift $z$. On linear scales, the quasar overdensity field $\delta_q(k,z)$ is connected to the underlying matter perturbation $\delta_m(k,z)$ with a local bias, $\delta_q(k,z) = b(z) \delta_m(k,z)$, where $b(z)$ is the linear bias of the quasar sample at redshift $z$. In general, $b(z)$ could also be scale-dependent due to primordial non-Gaussianity~\citep[PNG,][]{Dalal2008}, which introduces an additional scale-dependent part whose amplitude is parameterized by $f_{\rm NL}$. This scale dependence for local PNG decreases rapidly ($\propto k^{-2}$) with scale, and given the latest constraint $f_{\rm NL} = -0.9\pm 5.1$ reported by~\citet{Planck2018PNG}, we should be allowed to assume a scale-independent bias. It is worth mentioning that our estimators of the angular power spectra (Section~\ref{subsec:cls_est} below) do not directly rely on these theoretical predictions, and a redshift dependent $b(z)$ model may only matter in the $C_\Gamma$ calibration, as discussed below Eq.~\ref{eq:gamma_cor}.

Later in our analysis it will be useful to have analytic expressions for statistical errors of $C_\ell^{\kappa q}$ and $C_\ell^{qq}$. Assuming $\kappa$ and $q$ to be Gaussian fields, the sample variance of $C_\ell^{\kappa q}$ can be approximated as
\begin{equation}
    \begin{split}
        \sigma^2(C_\ell^{\kappa q}) &= \frac{1}{(2\ell+1)f_{\rm sky}^{\kappa q}}\left[(C_\ell^{\kappa q})^2 + (C_\ell^{\kappa\kappa}+N_\ell^{\kappa\kappa})(C_\ell^{qq}+N_\ell^{qq}) \right] \\
        &= \frac{(r_\ell^2 + 1)}{(2\ell+1)f_{\rm sky}^{\kappa q}} (C_\ell^{\kappa\kappa}+N_\ell^{\kappa\kappa})(C_\ell^{qq}+N_\ell^{qq})\, ,
    \end{split}
\label{eq:clkq_err}
\end{equation}
where $r_\ell \equiv C_\ell^{\kappa q}/[(C_\ell^{\kappa\kappa}+N_\ell^{\kappa\kappa}) (C_\ell^{qq}+N_\ell^{qq})]^{1/2}$ is known as the cross-correlation coefficient, $N_\ell^{\kappa\kappa}$ is the lensing reconstruction noise, $N_\ell^{qq}$ is the shot noise and $f_{\rm sky}^{\kappa q}$ is the overlapped sky coverage fraction of the surveys. Similarly, we have the variance for $C_\ell^{qq}$,
\begin{equation}
    \sigma^2(C_\ell^{qq}) = \frac{2}{(2\ell+1)f_{\rm sky}^{q}}(C_\ell^{qq}+N_\ell^{qq})^2\, .
\label{eq:clqq_err}
\end{equation}
When the multipoles $\ell$'s are averaged into bandpowers $p$'s (as discussed below) weighted by inverse variance (i.e. minimum variance average of Gaussian random vectors assuming no covariances), the uncertainty for the binned signal is given by 
\begin{equation}
    \sigma(C_p) = \left[\sum_{\ell\in p}\sigma^{-2}(C_\ell)\right]^{-1/2} \, .
\end{equation}
These analytic uncertainties, which include the well known lensing reconstruction and shot noise, have been widely used in doing forecasts. So it would be useful to have them as references and compared to the statistical errors estimated with simulations and jackknife resampling.

\subsection{Estimators} \label{subsec:cls_est}

Due to the noise and computational complexity, it is neither necessary nor possible to estimate $C_\ell$ for each multipole. Thus we bin the multipoles into bandpowers, denoted with subscript $p$~\footnote{To be clear, we use $p$ in this subsection. But for simplicity and consistency with the theory, we still use subscript $\ell$ for the bandpowers in the following sections.}. Here we briefly describe the estimators we use for $C_p^{\kappa q}$ and $C_p^{qq}$.

We estimate $C_p^{\kappa q}$ with the Pseudo-$C_\ell$ (PCL) estimator,
\begin{equation}
    \hat{C}_p^{\kappa q} = \sum_{p'}\left[\mathcal{M}^{-1}\right]_{pp'}\hat{D}_{p'}^{\kappa q}\, ,
\label{eq:clkq_est}
\end{equation}
where $\mathcal{M}$ is the binned mode coupling matrix computed with the masks of the two fields and $\hat{D}_p^{\kappa q}$ is the binned cross-power spectrum of the masked full sky maps,
\begin{equation}
    \hat{D}_p^{\kappa q} = \sum_{\ell\in p}w_\ell\left(\frac{1}{2\ell+1} \sum_{m=-\ell}^{\ell}\kappa^*_{\ell m}q_{\ell m}\right)\, ,
\end{equation}
where $w_\ell$ is the normalized weight of each multipole inside the bin, $\kappa_{\ell m}$ and $q_{\ell m}$ are the harmonics of the masked (i.e. with pixel values set to $0$ if not covered) $\kappa$ and $q$ maps. We use the fast implementation \texttt{NAMASTER}~\footnote{\url{https://github.com/LSSTDESC/NaMaster}}~\citep{Alonso2019} to do the computation. For the maps used in this work, the results of this more complicated PCL estimator are consistent with the results given by the simpler version $\hat{C}_p^{\kappa q} \simeq \hat{D}_p^{\kappa q}/f^{\kappa q}_{\rm sky}$, where the couplings between modes due to the geometry of the masks are ignored.

For the cross correlation, the noise in the two maps from separate surveys are usually uncorrelated and only contributes to the statistical error without causing systematic bias. However, the situation is more complicated for the auto correlation because the noise may not be correlated to the signal but is obviously correlated with itself and hence can significantly bias the signal. Thus for the estimation of $C_p^{qq}$, instead of the PCL estimator, we use the optimal quadratic minimum variance (QMV) estimator which marginalizes over the noise \citep{Tegmark1997}. We denote the pixelated quasar overdensity map with an 1-D vector $\bm{x}$ and the corresponding covariance matrix with $\mathbf{C}$. Defining the quadratic vector
\begin{equation}
    \hat{Q}_{p} = \frac{1}{2}\bm{x}^{\dagger}\mathbf{C}^{-1}\frac{\partial \mathbf{C}}{\partial C_p}\mathbf{C}^{-1}\bm{x}\, ,
\end{equation}
and the Fisher matrix
\begin{equation}
    \mathcal{F}_{pp'} = \frac{1}{2}\mathrm{Tr}\left(\mathbf{C}^{-1}\frac{\partial \mathbf{C}}{\partial C_p}\mathbf{C}^{-1}\frac{\partial \mathbf{C}}{\partial C_{p'}}\right)\, ,
\end{equation}
the estimator can be constructed as
\begin{equation}
    \hat{C}_p^{qq} = \sum_{p'}\left[\mathcal{F}^{\ -1}\right]_{pp'}\hat{Q}_{p'}\, .
\label{eq:clqq_est}
\end{equation}
The shot noise is properly fitted and marginalized in the estimation. In this work, $\bm{x}$ includes $\sim 10^6$ pixels, which makes it computationally impossible to invert $\mathbf{C}$ directly. We use the \textit{conjugate gradient method} to iteratively evaluate $\mathbf{C}^{-1}\bm{x}$ and the trace for the Fisher matrix. This optimal QMV estimator has been used in previous CMB and galaxy power spectra analysis, and we refer our readers to the references~\citep{Padmanabhan2001, Padmanabhan2003, Padmanabhan2007, Hirata2004, Hirata2008, Ho2008} for more details.

\subsection{Systematics \& calibrations}
Compared with the theoretical predictions in Eq.~\ref{eq:clkq_gr} and~\ref{eq:clqq}, the estimated power spectra can be biased due to several aspects, most of which are hard to be corrected in the estimators above and hence extra calibrations might be needed. For the quasars, the observed flux and measured redshift from the photometric and spectroscopic surveys are distorted due to the foreground density perturbation and RSD. There can also be bias due to redshift smearing. For the CMB survey, the temperature map can be contaminated by foregrounds, e.g. dust emission and point sources. Here we mainly focus on the bias caused by the distortion of quasar catalogs. The possible systematics due to contamination in CMB are discussed in Appendix~\ref{sec:cmb-sys}, where it is shown that the bias to $\hat{C}_\ell^{\kappa q}$ is negligible.

Our observed targets are distorted by the gravitational lensing of the foreground density perturbations. First, compared with the intrinsic value, the flux of an individual source can be either increased or decreased by lensing. This can cause bias to the flux or magnitude based target selection of the clustering catalog. Also, the observed angular distribution of the targets can be magnified. These effects can be quantified by the magnification bias $s$~\citep{Liu2014,Hui2007}, which can be measured with the slope of the cumulative apparent magnitude function,
\begin{equation}
    s = \left.\frac{d\log_{10}n_q(m<m_*)}{dm}\right|_{m=m_*}\, ,
\label{eq:mag_bias}
\end{equation}
where $m_*$ is the faint magnitude limit of the survey and $n_q(m<m_*)$ is the number of quasars that are apparently brighter than the survey limit. Depending on the value of $s$ and the linear bias $b$, our estimates of the power spectra can be more or less biased~\citep{Dizgah2016}. Following~\cite{Yang2018} (see Section 2 and Appendix A therein for the expressions using Limber approximation), we do the calibration by adding correction terms $\Delta\hat{C}_\ell^{\kappa q}$ and $\Delta\hat{C}_\ell^{qq}$ to our estimates from Eq.~\ref{eq:clkq_est} and Eq.~\ref{eq:clqq_est}. Besides $s$ and $b$, these corrections also depend on the measured CMB lensing auto-power spectrum $\hat{C}_\ell^{\kappa\kappa}$. The target selection for eBOSS quasars includes the magnitude cutoff for two frequency bands, $g<22$ or $r<22$~\citep{Myers2015}. Looking into the apparent point spread function (PSF) magnitudes, we find that the overall cutoff is mainly dominated by the $r$ band. With Eq.~\ref{eq:mag_bias}, we get $s\simeq 0.1$ for both NGC and SGC. The corrections are around $13\%$ for $C_\ell^{\kappa q}$ and $7\%$ for $C_\ell^{qq}$, which results in about $5\%$ calibrations in $E_G(\ell)$.

RSD describes the distortion in the observed radial positions of the targets due to peculiar velocities. Although the redshift details are mostly erased in the projection of 2D angular maps and 3D clustering is usually used in RSD analysis (as discussed in the next section), RSD could still bias the angular power spectra mainly due to the flow at the cutoff boundaries of the redshift bin. Starting from the additional RSD component in the window function ~\citep[see Eq. 27 in][]{Padmanabhan2007} and using Limber approximation, the bias on $C_\ell^{\kappa q}$ due to RSD can be described with a higher-order term,
\begin{equation}
\begin{split}
    C_\ell^{\kappa q, r} = \frac{3\Omega_m^0H_0^2}{2c^2} \int_{z_1}^{z_2}\frac{cdz}{H(z)} K(\ell,\chi) f(z) \chi^{-2} W(z)
    P_m\left(\frac{\ell+1/2}{\chi},z\right)\, ,
\end{split}
\label{eq:clkq_r}
\end{equation}
where $f(z)$ is the linear growth rate and
\begin{equation}
\begin{split}
    K(\ell, \chi) \equiv\ &\frac{2\ell^2+2\ell-1}{(2\ell-1)(2\ell+3)}\phi(\chi) \\
    &- \frac{\ell(\ell-1)}{\sqrt{2\ell-3}(2\ell-1)\sqrt{2\ell+1}}\phi\left(\frac{\ell-3/2}{\ell+1/2}\chi\right) \\
    &- \frac{(\ell+1)(\ell+2)}{\sqrt{2\ell+1}(2\ell+3)\sqrt{2\ell+5}}\phi\left(\frac{\ell+5/2}{\ell+1/2}\chi\right)\,,
\end{split}
\end{equation}
where $\phi(\chi)=f_q(z)\frac{H(z)}{c}$ is the normalized quasar redshift distribution as a function of the comoving distance.
Similarly, the bias on $C_\ell^{qq}$ can be described with two extra terms,
\begin{equation}
    C_\ell^{qq,r} = \int_{z_1}^{z_2}\frac{cdz}{H(z)} K(\ell,\chi) \frac{H(z)}{c}\chi^{-2}f(z)b(z)P_m\left(\frac{\ell+1/2}{\chi},z\right)
\label{eq:clqq_r}
\end{equation}
and
\begin{equation}
    C_\ell^{qq,rr} = \int_{z_1}^{z_2}\frac{cdz}{H(z)} K^2(\ell,\chi) \frac{H(z)}{c} \chi^{-2} f^2(z) P_m\left(\frac{\ell+1/2}{\chi},z\right)\, .
\label{eq:clqq_rr}
\end{equation}
For the multipoles we are considering, $100\leq \ell \leq 1000$, we find $C_\ell^{\kappa q, r}/C_\ell^{\kappa q}<10^{-4}$, $C_\ell^{qq,r}/C_\ell^{qq}<10^{-4}$ and $C_\ell^{qq,rr}/C_\ell^{qq}<10^{-8}$, where $C_\ell^{\kappa q}$ and $C_\ell^{qq}$ are the true power spectra in Eq.~\ref{eq:clkq} and Eq.~\ref{eq:clqq}. So the bias due to RSD is completely negligible. This is expected since the redshift bin $0.8 < z < 2.2$ for the quasar sample is wide, while the distortion only happens around the edges of the bin. Similarly, the bias due to redshift smearing error from the redshift fitting pipeline should also be negligible for these angular power spectra.

\section{Redshift-Space Distortion} \label{sec:rsd}
We estimate the RSD parameter $\beta$ of the quasar sample at the effective redshift by fitting an analytic model to the monopole and quadrupole of the configuration space 2-point correlation function (2PCF).

\subsection{Two point correlation function}
The 2PCF is estimated with the standard Landy \& Szalay estimator~\citep{Landy1993},
\begin{equation}
    \xi = \frac{\langle DD \rangle - 2\langle DR \rangle + \langle RR \rangle}{\langle RR \rangle}\, ,
\end{equation}
where $D$ is the data catalog, $R$ is the random catalog and $\langle \ \  \rangle$ denotes the normalized pair count between two catalogs. We sample the pair counts in $(s,\mu)$ bins, where $s$ is the separation distance and $\mu$ is the cosine of the angle between the line-of-sight (LOS) and separation vectors. We use a bin size of $5\Mpcph$ for $s$ and $0.01$ for $\mu$. The multipoles are extracted by expanding the 2PCF in Legendre polynomials,
\begin{equation}
    \xi(s,\mu) = \sum_\ell \xi_\ell(s) L_\ell(\mu)\, ,
\end{equation}
where
\begin{equation}
    \xi_{\ell}(s) = \frac{2\ell+1}{2} \int_{-1}^1 \xi(\mu,s) L_{\ell}(\mu) d\mu \, .
\end{equation}
The two lowest order even multipoles, monopole $\xi_0(s)$ and quadrupole $\xi_2(s)$, are used in the following fitting process. The non-zero quadrupole results from the peculiar velocity due to gravity and contains the information about the growth of the structure and RSD. For jackknife resampling, the pair count process is optimized to get the results for all samples in one run~\footnote{\url{https://gitlab.com/shadaba/CorrelationFunction}}. For the large number of mocks, we count the pairs with \texttt{Corrfunc}~\footnote{\url{https://github.com/manodeep/Corrfunc}}~\citep{Sinha2020}.

\subsection{CLPT-GS model}
The analytic model of the 2PCF we use is a combination~\citep{Wang2013} of the Convolution Lagrangian Perturbation Theory (CLPT)~\citep{Carlson2012} and the Gaussian Streaming (GS) model~\citep{Reid2011}.
In the GS model, the correlation function in redshift space is given by
\begin{equation}
    1 + \xi(s,\mu) = \int dy\frac{1+\xi(r)}{\sqrt{2\upi\sigma_{12}^2(r,\mu)}}\exp\left\{-\frac{\left[s\mu - y - \mu v_{12}(r)\right]^2}{2\sigma_{12}^2(r, \mu)}\right\} \, ,
\end{equation}
where the real space correlation function $\xi(r)$, pairwise velocity $v_{12}(r)$ and velocity dispersion $\sigma_{12}^2(r,\mu)$ are outputs from CLPT modified by the growth rate $f$ and the first and second-order Lagrangian bias, $F'$ and $F''$. On linear scales, the (Eulerian) bias and RSD parameter are given by $b=1+F'$ and $\beta=f/b$ respectively. $F'$ and $F''$ can also be constrained with a single overdensity parameter $\nu$ through peak-background split~\citep{White2014},
\begin{equation}
\begin{split}
    F' &= \frac{1}{\delta_c}\left[ a\nu^2 -1 + \frac{2p}{1+(a\nu^2)^p} \right]\,,\\
    F'' &= \frac{1}{\delta_c^2}\left[ a^2\nu^4 - 3a\nu^2 + \frac{2p(2a\nu^2+2p-1)}{1+(a\nu^2)^p} \right]\,,
\end{split}
\label{eq:pbs}
\end{equation}
where $a=0.707$ and $p=0.3$ with the Sheth-Tormen mass function~\citep{Sheth1999}, and $\delta_c=1.686$ is the linear critical overdensity of spherical collapse. To account for the finger-of-god (FoG) effect and redshift smearing error, $\sigma_{12}^2$ is modified by adding a nuisance term $\sigma_{\text{tot}}^2 = \sigma_{\text{FoG}}^2 + \sigma_z^2$, where $\sigma_{\text{FoG}}$ and $\sigma_z$ are degenerate in this model. CLPT takes the matter power spectrum as a input, which is calculated using \texttt{CAMB}~\footnote{\url{https://camb.info}}~\citep{Lewis2000} with our fiducial cosmological parameters.
This CLPT-GS model has been used in the RSD analysis of CMASS galaxies in $0.43 < z < 0.7$~\citep{Alam2015}, BOSS DR12 galaxies in $0.2 < z < 0.7$~\citep{Satpathy2017}, and eBOSS DR14 quasars in $0.8 < z < 2.2$~\citep{Zarrouk2018}.

\subsection{Parameters distribution sampling}

\begin{figure}
    \centering
    \includegraphics[width=\columnwidth]{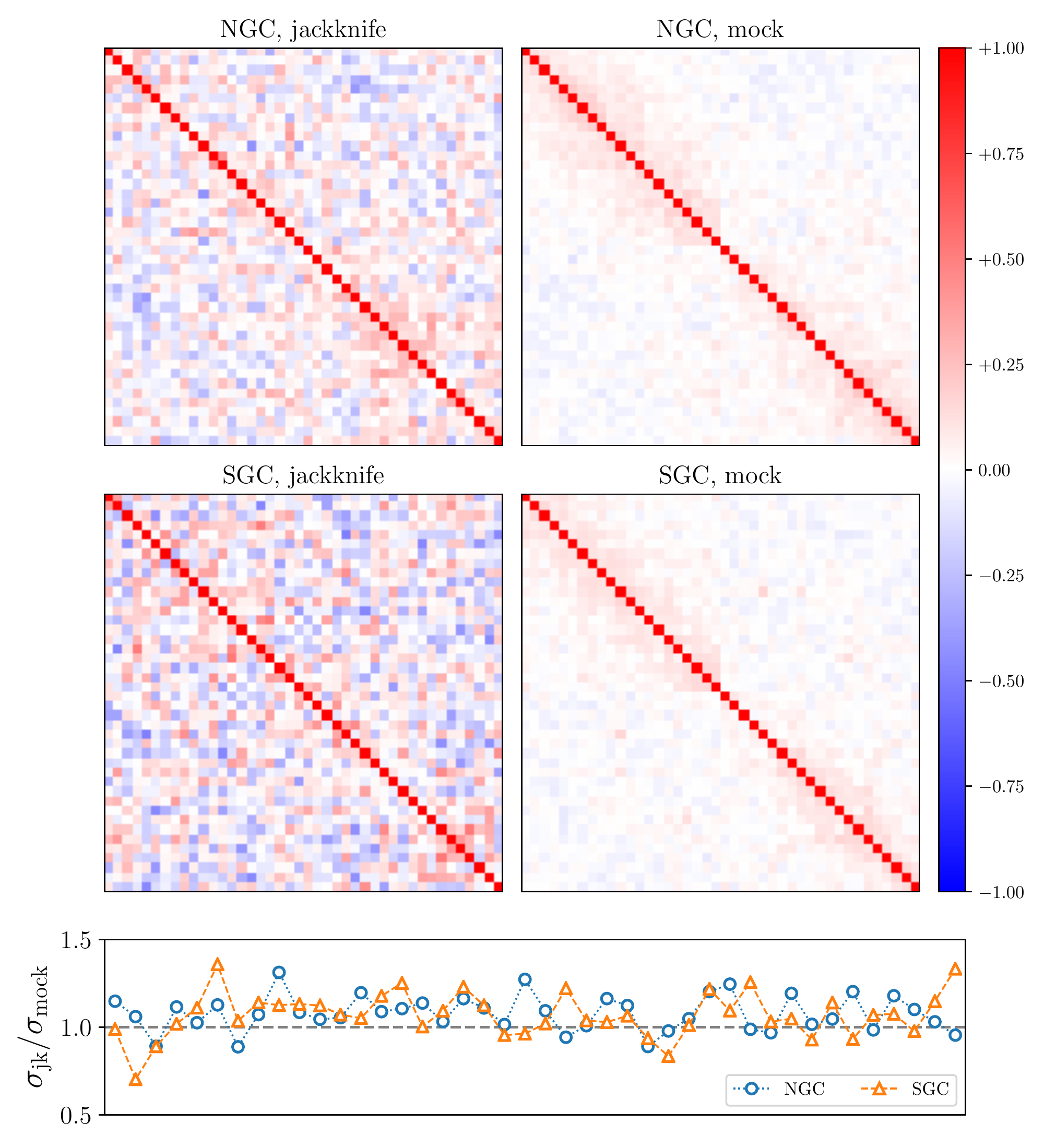}
    \caption{Correlation matrices (Eq.~\ref{eq:cor}) of $\xi_{0,2}$, a 1-D vector consists of monopole and quadrupole, estimated with jackknife resampling (\textit{left}) and $1000$ EZ mocks (\textit{right}) for NGC (\textit{upper}) and SGC (\textit{middle}). $\xi_0$ and $\xi_2$ sample points with separation distances $30\leq s\leq 135$ are included (Fig.~\ref{fig:xi02}), which gives $42$ data points in total. The \textit{lower} panel shows the ratio of the $1\,\sigma$ errors estimated with the two methods, i.e. $\sigma_{\rm jackknife}/\sigma_{\rm mock}$.}
    \label{fig:xi02_cov}
\end{figure}

Given the data and analytic model of the 2PCF, we construct a multivariate Gaussian likelihood function,
\begin{equation}
    \mathcal{L}(\theta|\hat{\xi}) \propto \exp\left[-\frac{1}{2}\left(\xi(\theta)-\hat{\xi}\right)^{T}\hat{\mathbf{C}}^{-1}\left(\xi(\theta)-\hat{\xi}\right)\right]\, ,
\label{eq:rsd_likelihood}
\end{equation}
where $\hat{\xi}$ is the data vector consists of $\hat{\xi}_0$ and $\hat{\xi}_2$, $\hat{\mathbf{C}}$ is the estimated covariance matrix of $\xi$, and $\xi(\theta)$ is the output of the CLPT-GS model described above with the set of free parameters denoted as $\theta$. We estimate $\hat{\mathbf{C}}$ with $1000$ EZ mocks and do the correction as described in Section~\ref{subsec:cov}. As a comparison, $\hat{\mathbf{C}}$ is also estimated with jackknife resampling. We plot the correlation matrices of $\hat{\mathbf{C}}$ with both methods and the ratio of the statistical errors in Fig.~\ref{fig:xi02_cov}.
We can see that compared with the mocks, jackknife resampling tends to overestimate the statistical errors and the relative strength of the covariances (i.e. the off-diagonal terms). Considering that the number of jackknives we are using is not very large, which hence may not be able to give us well-constrained estimates, here we use the simulated $\hat{\mathbf{C}}$ in the likelihood function.
The set of free parameters $\theta$ includes the RSD parameter $\beta$, the overdensity parameter $\nu$ and a nuisance velocity dispersion term $\sigma_{\rm tot}$. Flat priors are used for these parameters and the posterior distribution is sampled using Markov Chain Monte Carlo (MCMC) with \texttt{emcee}~\footnote{\url{https://github.com/dfm/emcee}}~\citep{Foreman-Mackey2013}.

\section{Results} \label{sec:results}
In this section, we first discuss the methods of combining NGC and SGC. Then we present our estimates of the angular power spectra $C_\ell^{\kappa q}$ and $C_\ell^{qq}$, and the RSD parameter $\beta$. These are then combined into $E_G(\ell)$ at the $5$ bandpowers, with which we find the best-fit scale-independent $\bar{E}_G$ estimate.

\subsection{Combination of NGC and SGC} \label{subsec:combine_ns}
As mentioned in Section~\ref{subsec:quasar_catalogs}, the quasar sample comprises two catalogs, which correspond to two separate regions on the sky, namely NGC and SGC. A proper combination of the two caps, which we denote as NS, should give us better constrained estimates. Throughout the data analysis pipeline in this work, this process can be conducted at several stages.

First, at the raw data level, the simplest approach is to put the two caps together before doing any estimation. For the quasar overdensity map, we may simply use all the quasars in the two catalogs to make one map or merge the two overdensity maps into one. For the estimation of the correlation function, we may combine the pair counts in the Landy \& Szalay estimator. However, we do not do the combination at this data level since NGC and SGC are observed with different photometric calibrations and have different number densities (Fig.~\ref{fig:zdist}), which result in different shot noise and other possible systematics. For the estimation of $C_\ell^{qq}$, where the shot noise contributes much more than the signal at smaller scales, this simple combination of two maps with different shot noise is not optimal.

Instead of combining the data of the two caps directly, we measure $C_\ell$'s and $\beta$ separately for the two caps, which are then averaged to get the estimates for NS. This process is conducted for the full data sample, simulations and jackknife samples. Assuming no cross correlation between the two caps, the average is weighted with inverse variances, which are estimated with the $300$ simulations. For jackknife resampling, $91=56+35$ jackknife estimates for NS are constructed by averaging each of the jackknife estimates from one cap with the full estimate from the other cap. It is worth mentioning that the jackknife estimates for NS are not constructed by simply stacking NGC and SGC estimates together, since the $91$ jackknives should make up a complete sample from which one jackknife region is left out each time. This also requires that we are using equal weights when making jackknife regions for NGC and SGC separately in order to make sure that they are statistically equivalent. Adhering to the advantage of being dependent only on the data, the variances used in these averages are also estimates from jackknife resampling instead of simulations or analytic uncertainties. These variances for jackknives should also be rescaled with the ratio of $f_{\rm sky}$'s between the \textit{leave-one-out} jackknife mask and the full mask, while the difference is negligible.

At last, we may also do the average with estimates of $E_G(\ell)$ or $\bar{E}_G$ for the two caps. As long as the error distributions of $C_\ell$'s, $\beta$ and $E_G$ are approximately Gaussian, this should be consistent with the method above.

\subsection{\texorpdfstring{$C_\ell^{\kappa q}$ and $C_\ell^{qq}$}{Clkq and Clqq}} \label{subsec:cls_res}

We consider the multipoles $100\leq \ell \leq 1000$ for our analysis of the angular power spectra and $E_G(\ell)$. This corresponds to the linear scales $19 < \chi_\perp < 190\Mpcph$ with the radial comoving distance $\chi(z=1.5)=3029\Mpcph$ given our fiducial cosmology. We do not consider smaller scales since $E_G$ is well defined only on linear scales. The largest scale that can be probed is limited by the spatial size of the sample and the trade-off between the number of jackknives. These multipoles are binned into $5$ evenly spaced bandpowers on the log scale, and thus $\ell$ indices for the estimates denote the bandpowers. We do not use narrower bins because the low SNR of $C_\ell^{qq}$ bandpowers at small scales could result in outliers in the $E_G(\ell)$ estimates with the $300$ simulations, whose error distribution would no longer be appropriate for estimating the Gaussian covariance matrix. Also, $E_G$ as a ratio of noisy quantities can be biased, hence using wider bins with smaller errors is preferable.

\begin{figure}
    \centering
    \includegraphics[width=\columnwidth]{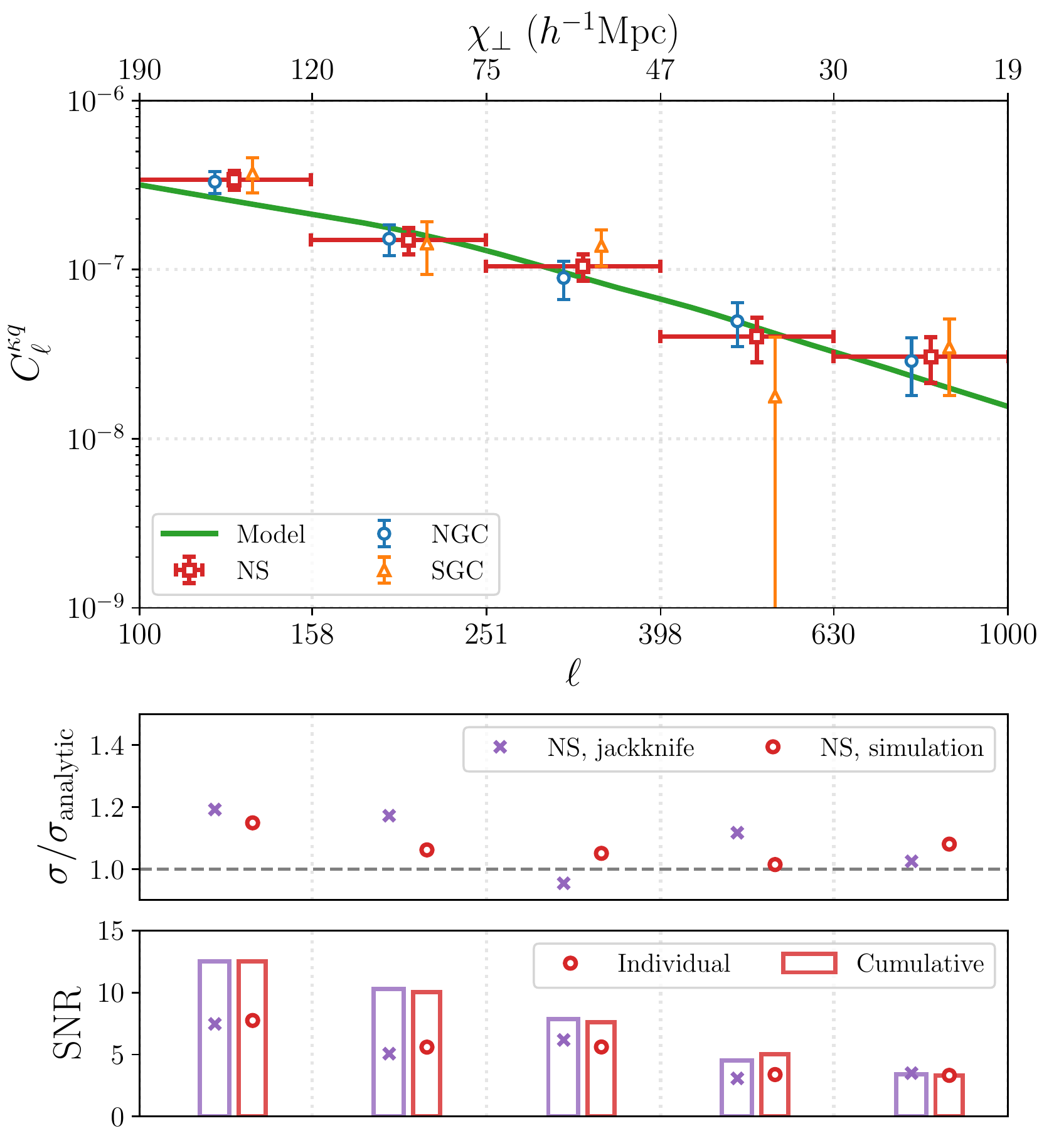}
    \caption{CMB lensing convergence $\kappa$ $\times$ quasar overdensity $q$ angular cross-power spectra. The crosses are estimates using \textit{Planck} 2018 CMB lensing map and eBOSS DR16 quasar clustering catalogs. We shifted the data points of NGC and SGC horizontally in the plot for reading convenience. For reference, we also plot the analytic model (Eq.~\ref{eq:clkq_gr}) with a linear bias $b=2.32$ fitted from $C_\ell^{qq}/C_\ell^{\kappa q}$. The statistical $1\,\sigma$ errors are estimated with $300$ simulations. The \textit{middle} panel includes the comparison of error estimates for NS using simulations and jackknife resampling with the analytic uncertainties (Eq.~\ref{eq:clkq_err}). Individual and cumulative SNRs for NS over the bandpowers are shown in the \textit{lower} panel, where the cumulative SNR starts from the smallest scale (i.e. highest $\ell$) and the covariances between scales are included (Eq.~\ref{eq:snr}).}
    \label{fig:clkq}
\end{figure}

\begin{figure}
    \centering
    \includegraphics[width=\columnwidth]{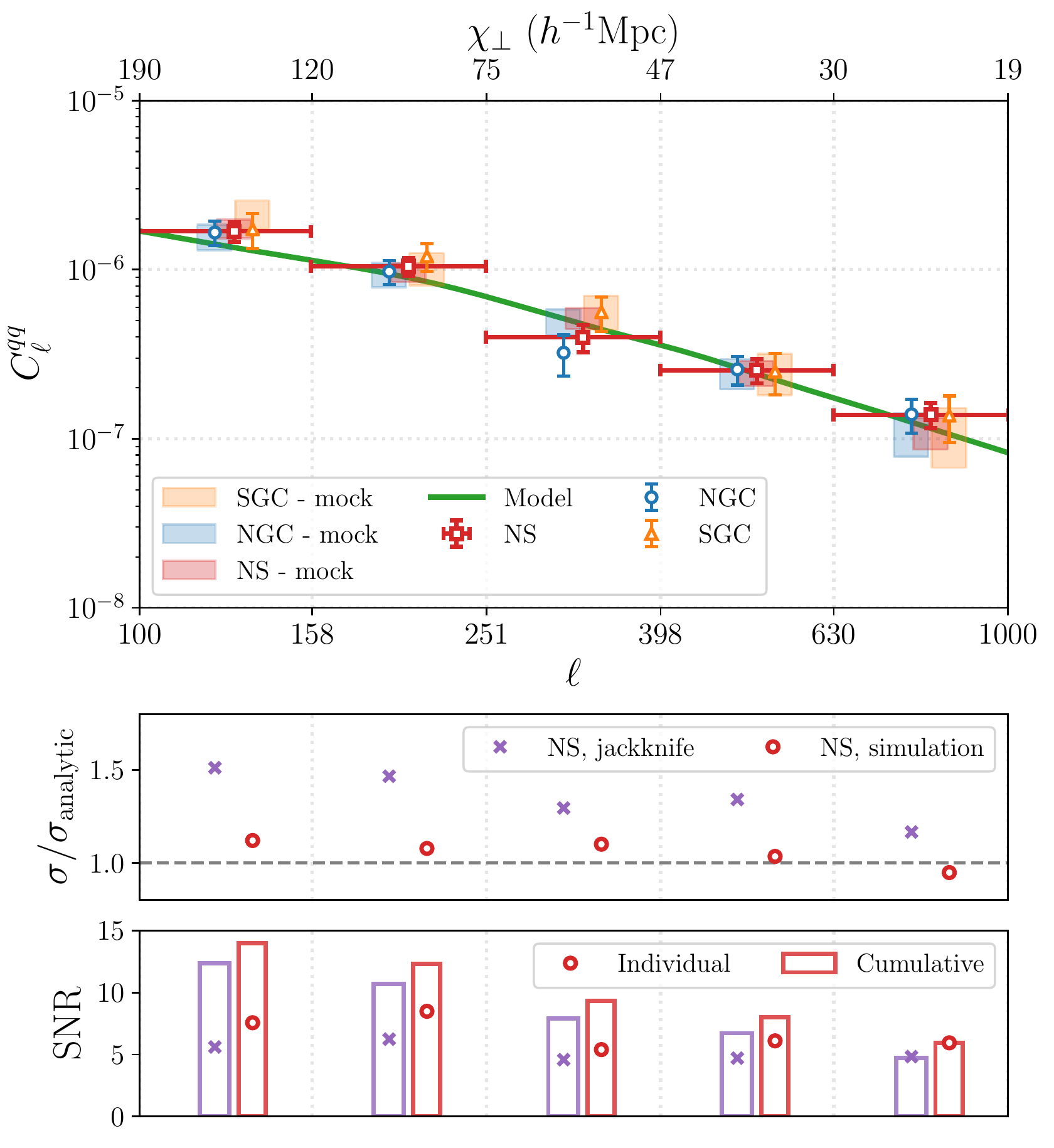}
    \caption{Quasar overdensity angular auto-power spectra, with similar information as Fig.~\ref{fig:clkq}. The shaded area denotes the average and $1\,\sigma$ error bar of estimates from $300$ EZ mocks. The analytic model in Eq.~\ref{eq:clqq} is plotted for reference, with the same bias used in Fig.~\ref{fig:clkq}. The analytic uncertainty is computed with Eq.~\ref{eq:clqq_err}.}
    \label{fig:clqq}
\end{figure}

The estimates of $C_\ell^{\kappa q}$ and $C_\ell^{qq}$ are shown in the \textit{upper} panels in Fig.~\ref{fig:clkq} and Fig.~\ref{fig:clqq} respectively, with the statistical $1\,\sigma$ errors given by simulations.
For reference, we also plot the analytic models discussed in Section~\ref{sec:cls-theory}, with a linear bias $b$ fitted from $C_\ell^{qq}/C_\ell^{\kappa q}$. It is worth noticing that for $C_\ell^{qq}$ estimates with quasar mocks, NGC and SGC are not very well consistent on the largest-scale bandpower. This might be caused by observational effects like completeness levels or systematic weights that are different for the two caps. A better understanding requires more simulations with different possible systematics applied, which we leave for future work. We do not have $C_\ell^{\kappa q}$ signals with simulations since as mentioned, our simulated $\kappa$ maps and quasar mocks are not correlated. Besides simulations, the statistical errors are also estimated with jackknife resampling. In the \textit{middle} panels, we show the comparison of the error estimates from both methods with the analytic uncertainties in Eq.~\ref{eq:clkq_err} and~\ref{eq:clqq_err}, where the quasar bias and shot noise are derived from data. As expected, the error estimates are mostly higher than the analytic uncertainties where only the lensing reconstruction and shot noise are considered. Even though our simulated $\kappa$ maps and quasar mocks are not correlated, the underestimation in $\sigma(C_\ell^{\kappa q})$ is negligible due to the low cross correlation coefficient $r_\ell<0.2$ (see Eq.~\ref{eq:clkq_err}).
We measure the marginalized SNR over scales with the full covariance matrix
\begin{equation}
    {\rm SNR}\left(C_\ell\right) = \left( \sum\nolimits_{\ell,\ell'} C_\ell \mathbf{C}^{-1}_{\ell\ell'} C_{\ell'} \right)^{1/2}
\label{eq:snr}
\end{equation}
to quantify the overall strength of the signal, where as mentioned above the summation runs over the $5$ bins for the estimates. The individual SNR for each bandpower and the cumulative SNRs starting from the highest-$\ell$ band are shown in the \textit{lower} panels. For $C_\ell^{\kappa q}$, both methods give similar errors and hence comparable SNRs. While for $C_\ell^{qq}$, jackknife resampling errors are higher than that from simulations. We are not doing any fittings with these angular power spectra, so more comparisons between the two methods are discussed in Section~\ref{subsec:eg_res}, where the covariance matrices for $E_G(\ell)$ are presented. With the simulated covariance matrices, we get overall ${\rm SNR}(C_\ell^{\kappa q})=12.5$ and ${\rm SNR}(C_\ell^{qq})=14.0$ for NS. Although the SNR for each band depends on our binning scheme, the overall value should remain roughly the same.

\subsection{RSD parameter}

\begin{figure}
    \centering
    \includegraphics[width=\columnwidth]{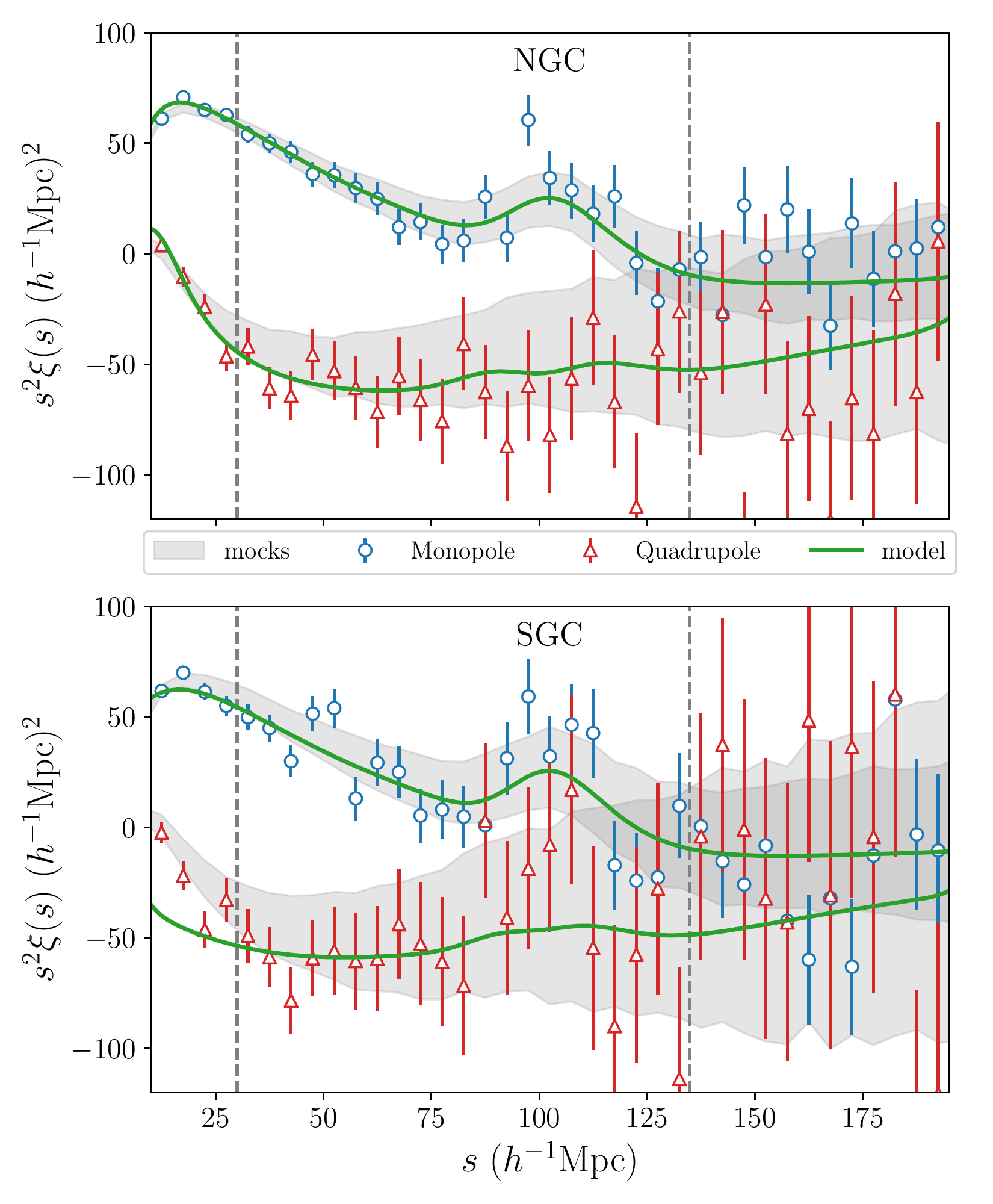}
    \caption{The monopole and quadrupole of the 2PCF with the best-fit CLPT-GS model. The crosses are estimates using eBOSS DR16 quasar NGC (\textit{upper}) and SGC (\textit{lower}) clustering catalogs. The grey shaded area denotes the mean and $1\,\sigma$ error of the $1000$ EZ mocks used to estimate the covariance matrix in Eq.~\ref{eq:rsd_likelihood}. Notice that the overall sky mask with CMB lensing has been applied on both the data and EZ mock catalogs. The two vertical dashed lines enclose the data points used in RSD fitting, with separation distances $30\leq s\leq 135\Mpcph$.}
    \label{fig:xi02}
\end{figure}
We show the estimated monopole and quadrupole of the 2PCF of the quasar catalogs in Fig.~\ref{fig:xi02}, along with the best-fit CLPT-GS model and the $1000$ EZ mocks. Data points with separation distances $30\leq s\leq 135\Mpcph$ are included in the RSD fitting. We do not use smaller scales $s<30\Mpcph$ where the CLPT-GS model has not been validated. We apply larger scale cutoff to optimize the model calculation~\citep{Alam2015} and remove any very large scale systematic in the QSO sample~\citep{Castorina2019}. Also, the contribution to our RSD fitting for the scale-independent parameters from these larger scales should be negligible due to the large errors. The goodness of fitting is given as $\chi^2/\text{dof} = 35/40\ (45/40)$ for NGC (SGC).

\begin{figure}
    \centering
    \includegraphics[width=\columnwidth]{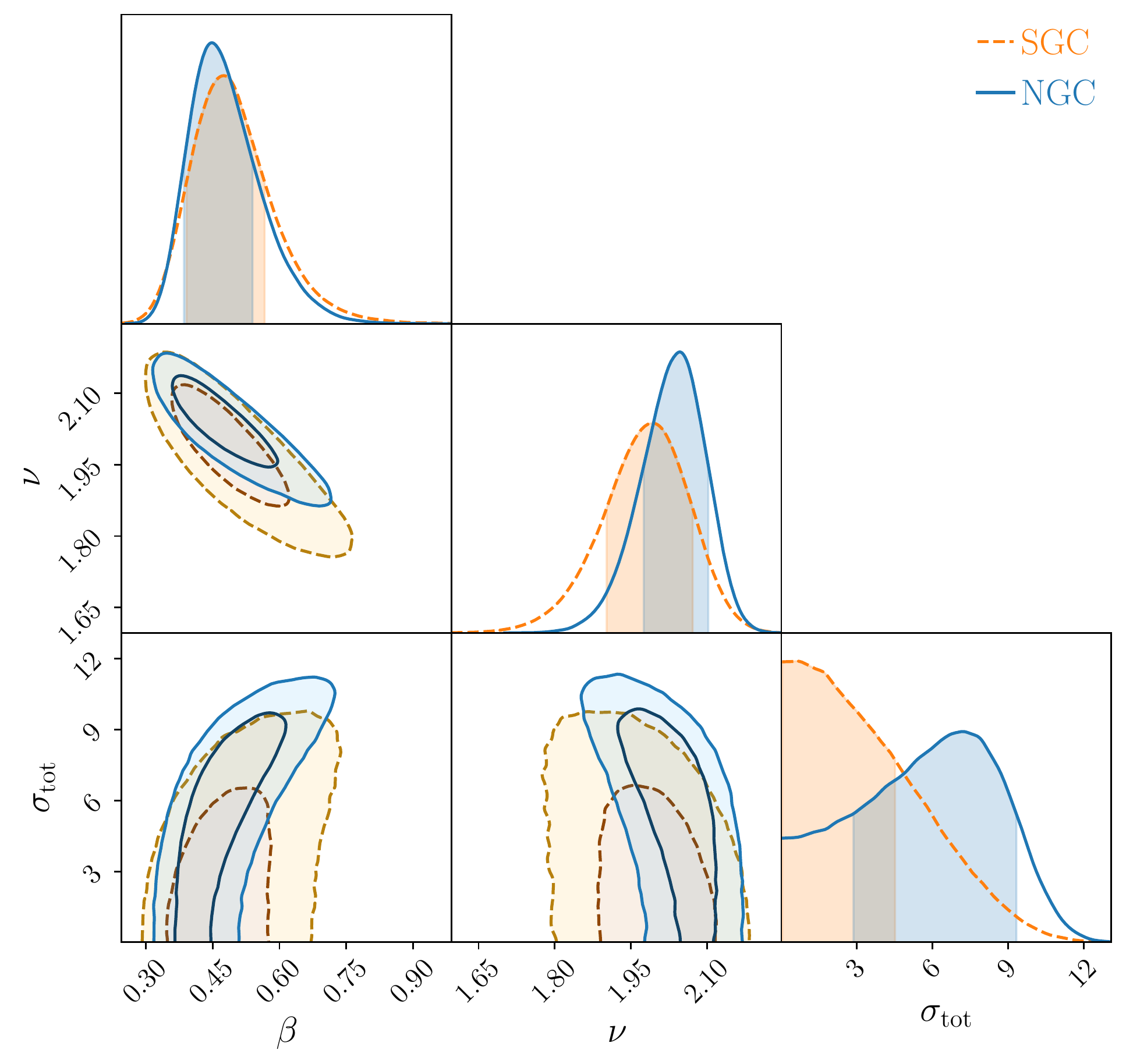}
    \caption{Posterior distributions of the parameters in RSD fitting, sampled with MCMC. The properties of the marginalized distributions of individual parameters are summarized in Table~\ref{tab:rsd_pars}.}
    \label{fig:rsd_pars}
\end{figure}

The posterior distributions with flat priors (i.e. likelihood functions) of the RSD parameter $\beta$, the overdensity parameter $\nu$ and the nuisance velocity dispersion parameter $\sigma_{\rm tot}$ are shown in Fig.~\ref{fig:rsd_pars}.
For $\beta$ and $\nu$, while slight skewness is observed, the distributions are approximately Gaussian around the maximum likelihood estimates. This skewness might be caused by the strong cross correlation with $\sigma_{\rm tot}$ at large values, as we can tell from the banana-shaped contours. These covariances with velocity dispersion might be better constrained with an optimized modelling that breaks the degeneracy between the FoG effect and the redshift smearing error. For $\sigma_{\rm FoG}$, a scale-dependent analytic model would be more accurate. The constraint on $\sigma_z$ could also be improved by constructing an informative prior based on redshifts measured with different methods. The best-fit estimates for the marginalized distribution of each parameter along with the confidence intervals are summarized in Table~\ref{tab:rsd_pars}.
\begin{table}
    \centering
    \caption{The maximum likelihood estimates of the RSD parameters with flat priors, where the $68.3\%$ confidence intervals are quoted with $\mathcal{L}(\theta_{-}) = \mathcal{L}(\theta_{+})$.}
    \label{tab:rsd_pars}
    \begin{tabular}{cccc}
        \hline
	    Parameter & $\beta$ & $\nu$ & $\sigma_{\rm tot}$ \\
	    Prior & $[0, 1]$ & $[1, 3]$ & $[0, 16]$ \\
	    \hline
	    NGC & $0.449^{+0.091}_{-0.063}$ & $2.047^{+0.056}_{-0.072}$ & $7.2^{+2.1}_{-4.3}$ \\
	    \hline
	    SGC & $0.474^{+0.093}_{-0.082}$ & $1.991^{+0.081}_{-0.089}$ & $0.63^{+3.89}_{-0.58}$ \\
	    \hline
    \end{tabular}
\end{table}
Though the confidence intervals inferred from posterior distributions are quoted for reference, these are not propagated to the error estimation of $E_G(\ell)$. As mentioned in Section~\ref{subsec:cov}, to estimate the full covariance matrix for $E_G(\ell)$, we also need to run all the simulations through the data analysis pipeline, including the RSD fitting process. For the 300 EZ mocks, the average along with the standard deviation of the best-fit estimates are $f\sigma_8 = 0.380 \pm 0.055$ for NGC and $f\sigma_8 = 0.366 \pm 0.067$ for SGC, which are consistent with the fiducial value $f\sigma_8 = 0.381$ given the cosmological parameters used in the EZ mock simulation.
The analysis of MCMC chains including the plots and statistics is conducted with the usage of \texttt{ChainConsumer}~\footnote{\url{https://github.com/samreay/ChainConsumer}}~\citep{Hinton2016}.

For our consistency test of $\Lambda$CDM-GR on the data, we are allowed to fix the fiducial cosmological parameters in this RSD fitting process since the \textit{Planck} 2018 results are measured to very high accuracy, and a flat prior based on this will not really change the marginalized distribution of the RSD parameters given the statistical accuracy.
If the true parameters are statistically different from \textit{Planck} results or $\Lambda$CDM-GR is not a proper model, we should be able to see the deviation of $E_G(\ell)$ estimates from the $\Lambda$CDM-GR prediction with \textit{Planck} parameters.
From $\beta$ and $\nu$, we can also infer the posterior distribution of the linear growth rate, which gives $f\sigma_8=0.424^{+0.064}_{-0.047}$ for NGC and $f\sigma_8=0.430^{+0.058}_{-0.057}$ for SGC. Our estimates are consistent with the eBOSS DR16 consensus result of the quasar sample, $f\sigma_8^{\bf{c}}(z_{\rm eff}=1.48)=0.462\pm 0.045$, which is a combination of the configuration space~\citep{Hou2020} and Fourier space~\citep{Neveux2020} analysis. The possible sources of difference include the overlapped mask with CMB lensing, fixed Alcock-Paczynski~\citep[AP;][]{Alcock1979} parameters and a different analytic model used in this work. The combination of $\xi(s)$ and $P(k)$ analysis could also help reduce the systematics in the consensus result~\citep{Smith2020}. A more detailed discussion of models and systematics in RSD fitting is out of the scope of this work, thus we refer our readers to the series of papers presenting the eBOSS final data release~\citep{eBOSS2020}.
In the RSD analysis of eBOSS DR14 quasar catalog using the same CLPT-GS model~\citep{Zarrouk2018}, a shift on the linear bias $\Delta b\sigma_8=0.037$ was observed when $F''$ was set free instead of fixed. So besides the main analysis using $\nu$ and peak-background split, we also do a test by running the RSD fitting with free $F'$ and $F''$ parameters on the data sample. For the RSD parameter we are interested in, we get $\beta=0.445^{+0.090}_{-0.059}$ for NGC and $\beta=0.458^{+0.101}_{-0.069}$ for SGC, which are consistent with the values in Table~\ref{tab:rsd_pars}.

\subsection{\texorpdfstring{$E_G$}{EG} estimates} \label{subsec:eg_res}

We combine our estimates of $C_\ell^{\kappa q}$, $C_\ell^{q q}$ and $\beta$ into $E_G(\ell)$ following Eq.~\ref{eq:eg_est}, with the calibration in Eq.~\ref{eq:gamma_cor} applied, which shifts the $E_G(\ell)$ signals lower for about $5\,\%$. We find the factor $\Gamma(\bar{z}=1.5) \simeq 0.74$ for both caps and NS.
\begin{figure}
    \centering
    \includegraphics[width=\columnwidth]{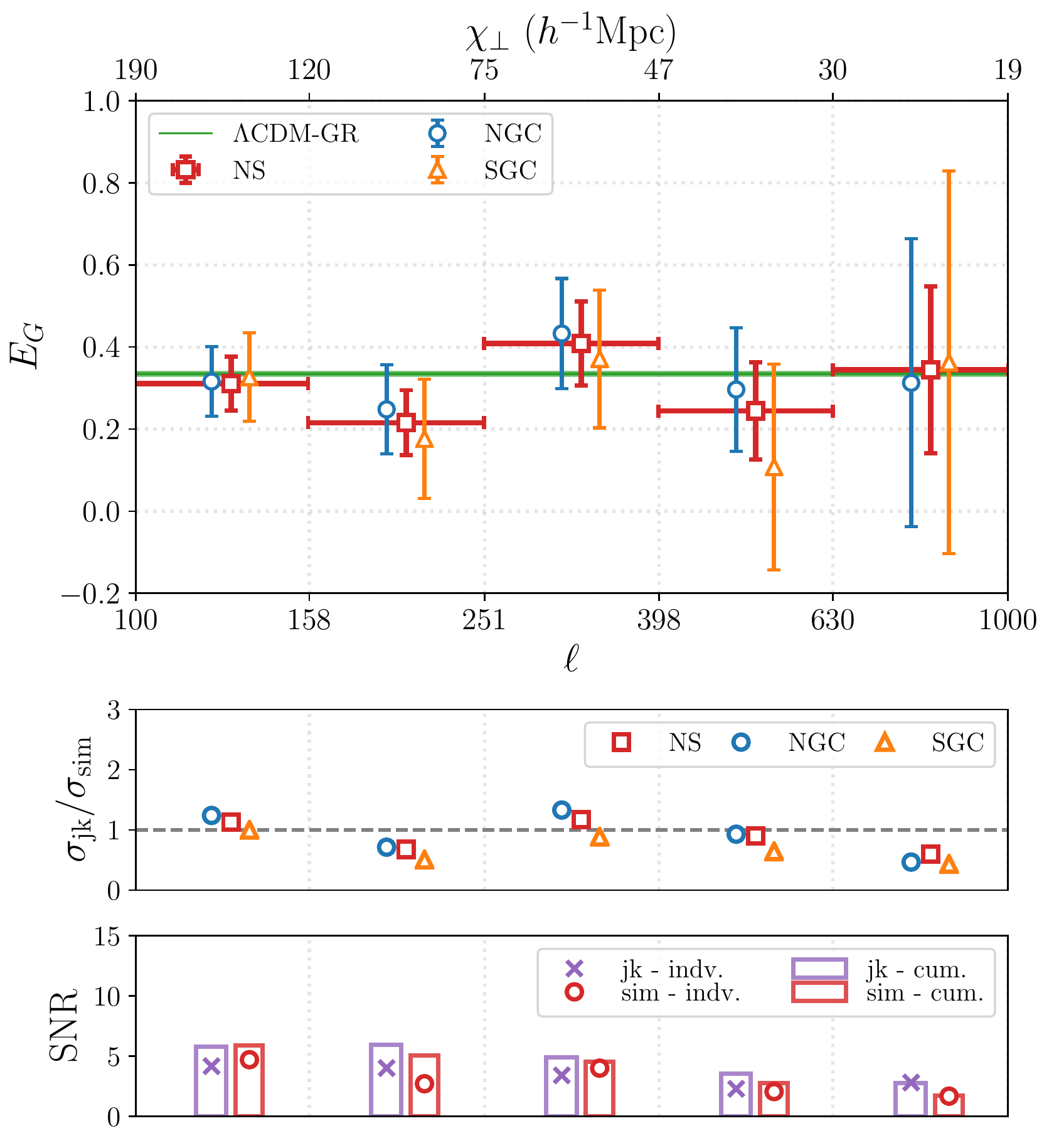}
    \caption{$E_G$ estimates using \textit{Planck} 2018 CMB lensing map and eBOSS DR16 quasar clustering catalogs. The data points for NGC and SGC are shifted horizontally in the plot for reading convenience. The green solid line and shaded area is the $\Lambda$CDM-GR prediction using the \textit{Planck} 2018 CMB+BAO matter density parameter and $1\,\sigma$ uncertainty, $\Omega_{\rm m} = 0.3111\pm 0.0056$. The $1\,\sigma$ error bars are estimated using simulations, with the comparison to errors given by jackknife resampling shown in the \textit{middle} panel as $\sigma_{\rm jackknife}/\sigma_{\rm simulation}$. As in Fig.~\ref{fig:clkq} and~\ref{fig:clqq}, the individual and cumulative SNRs of the bandpowers are shown in the \textit{lower} panel.}
    \label{fig:eg}
\end{figure}
The $E_G(\ell)$ estimates for the bandpowers are shown in Fig.~\ref{fig:eg}, where the $1\,\sigma$ statistical errors for individual bins are determined using simulations. These errors are also estimated using jackknife resampling, with the comparison shown in the \textit{middle} panel. We see that $E_G(\ell)$ estimates at all the $5$ bandpowers agree with the GR prediction at $1\,\sigma$ level, and we could not see an obvious scale-dependence pattern. Unlike the power spectra (Fig.~\ref{fig:clkq} and~\ref{fig:clqq}), the theoretical $E_G$ model does not depend on the clustering bias and is thus independent of the estimates, which makes the comparison between theory and observations more straightforward. Since the RSD parameter $\beta$ is assumed to be scale-independent, the fluctuations of $E_G(\ell)$ estimates are mainly determined by the ratio $C_\ell^{\kappa q}/C_\ell^{qq}$ (see Fig.~\ref{fig:rl}). As discussed in Section~\ref{subsec:combine_ns}, to get $E_G(\ell)$ estimates for NS, we can combine NGC and SGC at either the $\{C_\ell{\rm 's},\,\beta\}$ level or $E_G(\ell)$ level. The NS signals shown in Fig.~\ref{fig:eg} are derived using the first method, which are consistent with that using the second method. For the scale-averaged $\bar{E}_G$ discussed below, besides fitting $E_G(\ell)$ of NS, we can also do the fitting for NGC and SGC separately and then combine the results to get $\bar{E}_G$ for NS. We have tried all these methods, and the results are consistent within $3\%$, which is expected as for all the statistical quantities, the error distributions are approximately Gaussian and the two spatially separated caps should not be correlated for the scales we are considering.

Given the consistency between the $E_G$ estimates at all the $5$ scale bins and the scale-independent $\Lambda$CDM-GR prediction, we further improve the constraint on $E_G$ by fitting a constant $\bar{E}_G$ over the $5$ bins. We infer the best-fit value of $\bar{E}_G$ by maximizing the multivariate Gaussian likelihood function,
\begin{equation}
    \mathcal{L}(\bar{E}_G) \propto \exp\left\{-\frac{1}{2}\left[\hat{E}_G(\ell) - \bar{E}_G\right]^T\hat{\mathbf{C}}^{-1}\left[\hat{E}_G(\ell) - \bar{E}_G\right]\right\}\, ,
\label{eq:eg_likelihood}
\end{equation}
where $\hat{\mathbf{C}}$ is the estimated covariance matrix of $E_G(\ell)$. For this linear fitting model, the max-$\mathcal{L}$ point can be analytically written as
\begin{equation}
    \bar{E}_G = \frac{\sum_{\ell,\ell'}\hat{\mathbf{C}}_{\ell\ell'}^{-1}\hat{E}_G(\ell')}{\sum_{\ell,\ell'}\hat{\mathbf{C}}_{\ell\ell'}^{-1}}\, ,
\label{eq:eg_bestfit}
\end{equation}
with the statistical error
\begin{equation}
    \sigma\left(\bar{E}_G\right) = M \times \left(\sum\nolimits_{\ell,\ell'}\hat{\mathbf{C}}_{\ell\ell'}^{-1}\right)^{-1/2}\, ,
\label{eq:gaussian_err}
\end{equation}
where the summation runs over the $5$ bandpowers for the estimates, $\hat{\mathbf{C}}_{\ell\ell'}^{-1}$ is the $\ell,\ell'$ element of $\hat{\mathbf{C}}$ inverse with the correction in Eq.~\ref{eq:cinv} applied, and $M$ is the calibration factor in Eq.~\ref{eq:mab}.
As discussed in Section~\ref{subsec:cov}, we estimate $\hat{\mathbf{C}}$ with both jackknife resampling and simulations. One defect of the simulations is that the $\kappa$ maps and quasar mocks are not correlated. While the impact on $\sigma(C_\ell^{\kappa q})$ is negligible compared with the current lensing reconstruction and shot noise (see Section~\ref{subsec:cls_res}), the $C_\ell^{\kappa q}$ signal does matter in the covariance matrix of $E_G(\ell)$. To fix this issue, we shift the center of the error distribution of the $300$ simulated $C_\ell^{\kappa q}$'s from zero to the expected signal with a fiducial quasar bias measured from the data. By doing this, the distribution of the simulated $C_\ell^{\kappa q}$'s should be roughly equivalent to what we would get if the simulations were correlated.
\begin{figure}
    \centering
    \includegraphics[width=\columnwidth]{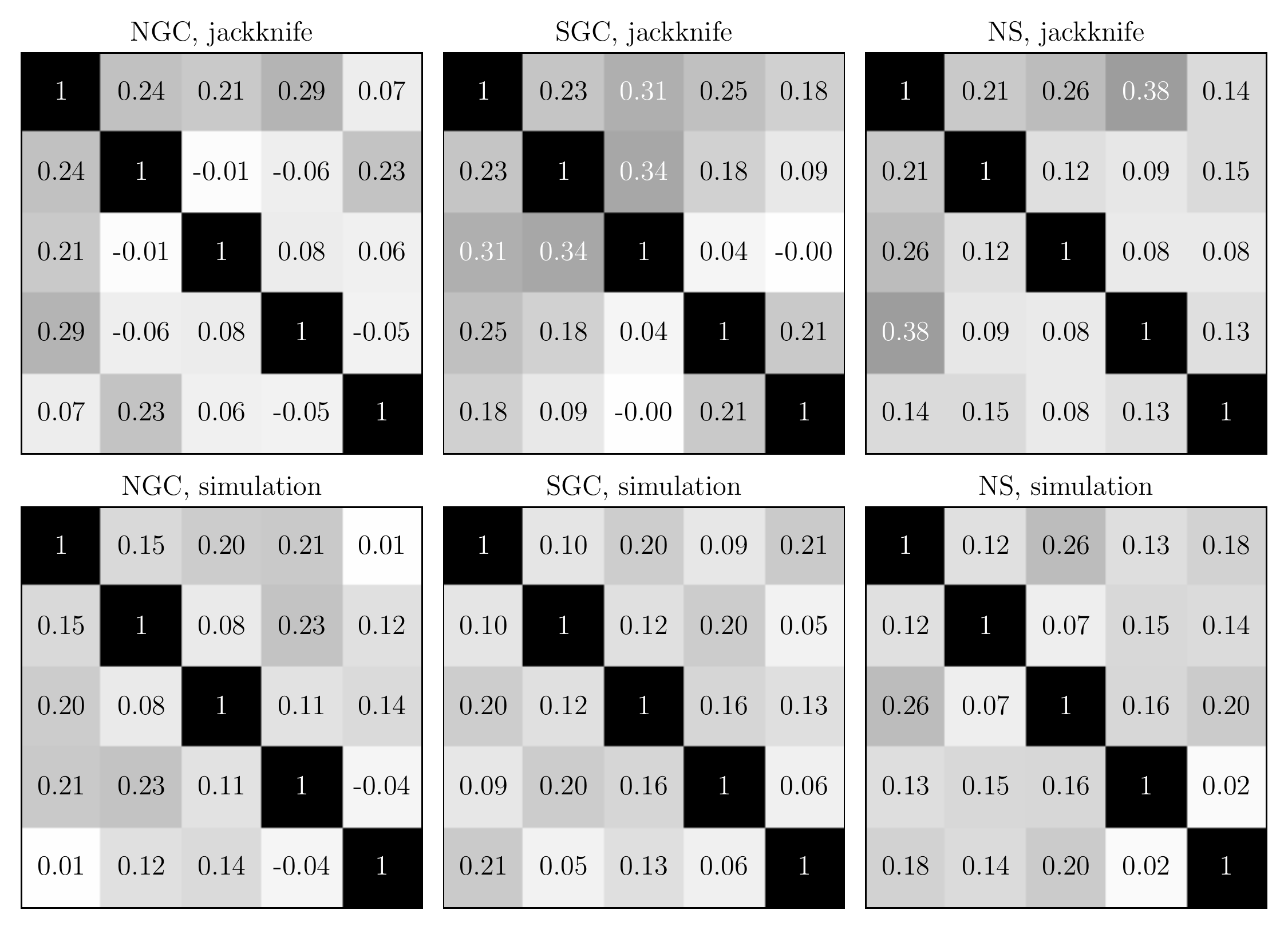}
    \caption{Estimated correlation matrices (Eq.~\ref{eq:cor}) of $E_G(\ell)$ with jackknife resampling (\textit{upper}) and $300$ simulations (\textit{lower}) for NGC, SGC and the combination, NS. The number of jackknife samples is $56$ for NGC, $35$ for SGC and $91$ for NS.}
    \label{fig:eg_cov}
\end{figure}
The correlation matrices (Eq.~\ref{eq:cor}) of $\hat{\mathbf{C}}$ from both methods are shown in Fig.~\ref{fig:eg_cov}, and the square root ratios of the diagonal terms are shown in the \textit{middle} panel in Fig.~\ref{fig:eg}.
We see that $\hat{\mathbf{C}}$'s given by both methods include non-negligible cross correlations between scales. This can be caused by the fact that we are using one scale-independent $\beta$ estimate for all bandpowers, which introduces the same variation for all of them and hence contributes to the covariances. To test if $\hat{\mathbf{C}}$'s given by jackknife or simulations are well constrained, we take another approach of estimating $\bar{E}_G$ by fitting the ratio $R_\ell\equiv C_\ell^{\kappa q}/C_\ell^{qq}$ over the 5 scale bins first, which is then divided by the scale-independent $\beta$. In that case, CM for $R_\ell$ instead of $E_G(\ell)$ is estimated but final $\bar{E}_G$ estimates should be consistent if CMs in both approaches are well constrained. More details are included in Appendix~\ref{sec:rl}. It is shown that the covariances of $R_\ell$ are much weaker (Fig.~\ref{fig:rl_cov}) than that of $E_G(\ell)$ (Fig.~\ref{fig:eg_cov}), which is expected without the same $\beta$ variation for all bins. The two approaches give consistent final results with $\hat{\mathbf{C}}$'s given by simulations. While with jackknife resampling, the final $\bar{E}_G$ estimates are more different, especially for SGC. One reason might be that the numbers of jackknives, with only $35$ samples for SGC, are not enough to get converged estimates. Also, for the two caps, the observational systematics in the imaging used to target quasars are different, which may result in different unknown bias. The poor constraint on $\hat{\mathbf{C}}$ for either or both of $E_G(\ell)$ and $R_\ell$ can then bias our fitting for $\bar{E}_G$.

\begin{table}
    \centering
    \caption{$E_G$ estimates at the effective redshift $\bar{z}=1.5$ averaged over scales $19\leq \chi_{\perp}\leq 190\Mpcph$ with \textit{Planck} 2018 CMB lensing map and eBOSS DR16 quasar clustering catalogs. Best-fit results for NGC, SGC and the combination NS with simulated $\hat{\mathbf{C}}$ are quoted with $1\,\sigma$ statistical errors. The deviations from $\Lambda$CDM-GR prediction $E_G(z=1.5)=0.3346$ with $\Omega_{m,0}=0.3111$ are also presented. The last row includes the best-fit estimates using $\hat{\mathbf{C}}$ from jackknife resampling, which are not reported as our final results due to the possible poor constraints on the covariance matrices (see text).}
    \label{tab:eg_ests}
    \begin{tabular}{cccc}
        \hline
        Cap & NS & NGC & SGC \\
        $E_G$ & $0.295\pm 0.054$ & $0.309\pm 0.068$ & $0.272\pm 0.087$ \\
        Deviation & $0.74\,\sigma$ & $0.38\,\sigma$ & $0.72\,\sigma$ \\\hline
        $E_G$ with $\hat{\mathbf{C}}_{\rm jk}$ & $0.253\pm 0.050$ & $0.283\pm 0.066$ & $0.214\pm 0.076$ \\\hline
    \end{tabular}
\end{table}
We summarize our best-fit estimates of the scale-averaged $\bar{E}_G$ in Table~\ref{tab:eg_ests}. Considering the result of the test above and the fact that the simulations we are using are designed to be as realistic as possible, i.e. including all the known systematics, we take the estimates with simulated $\hat{\mathbf{C}}$ as our primary results. Although the signals are different, the statistical errors given by the two methods are almost the same. We report a best-fit $\bar{E}_G(z\simeq 1.5)=0.295\pm 0.054$ estimate for NS, which is about $0.74\,\sigma$ lower than the $\Lambda$CDM-GR prediction with \textit{Planck} 2018 CMB+BAO $\Omega_m^0$. For the two separate caps, they agree with each other and NGC is more consistent with the GR prediction with a $0.38\,\sigma$ deviation.
\begin{figure}
    \centering
    \includegraphics[width=\columnwidth]{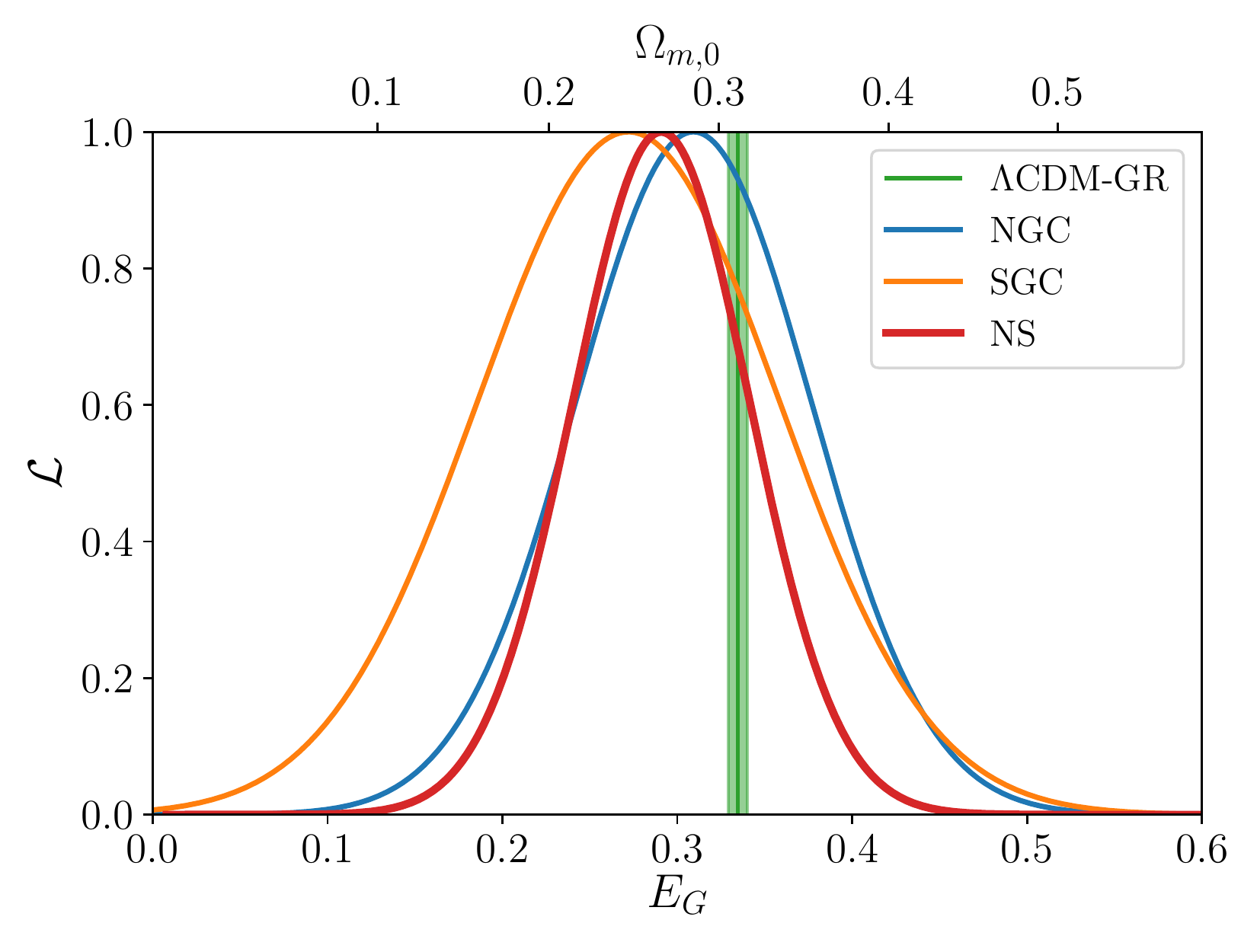}
    \caption{Likelihood functions of scale-averaged $E_G$, with covariance matrices estimated using simulations. The green line with shaded area corresponds to the $\Lambda$CDM-GR prediction with the \textit{Planck} 2018 CMB+BAO matter density and $1\,\sigma$ uncertainty, $\Omega_{m,0}=0.3111\pm 0.0056$.}
    \label{fig:eg_prob}
\end{figure}
For reference, the likelihood functions of $\bar{E}_G$ are shown in Fig.~\ref{fig:eg_prob}.

\section{Conclusions} \label{sec:conclusions}

$E_G$ is a promising probe of gravity on cosmological scales by combining gravitational lensing and LSS, with the advantage of being independent of the tracer bias and $\sigma_8$. In this work, we estimate $E_G$ at the effective redshift $z\sim 1.5$ over scales $19-190\Mpcph$ with the \textit{Planck} 2018 CMB lensing convergence map and SDSS eBOSS DR16 quasar clustering catalogs. This is the highest redshift and largest scale where $E_G$ has been estimated so far. We show that quasars are promising DM LSS tracers for both auto correlation clustering analysis and cross correlation with the weak gravitational lensing signal reconstructed from CMB. Our results are in line with the $\Lambda$CDM-GR prediction within $1\,\sigma$ confidence interval.
\begin{figure}
    \centering
    \includegraphics[width=\columnwidth]{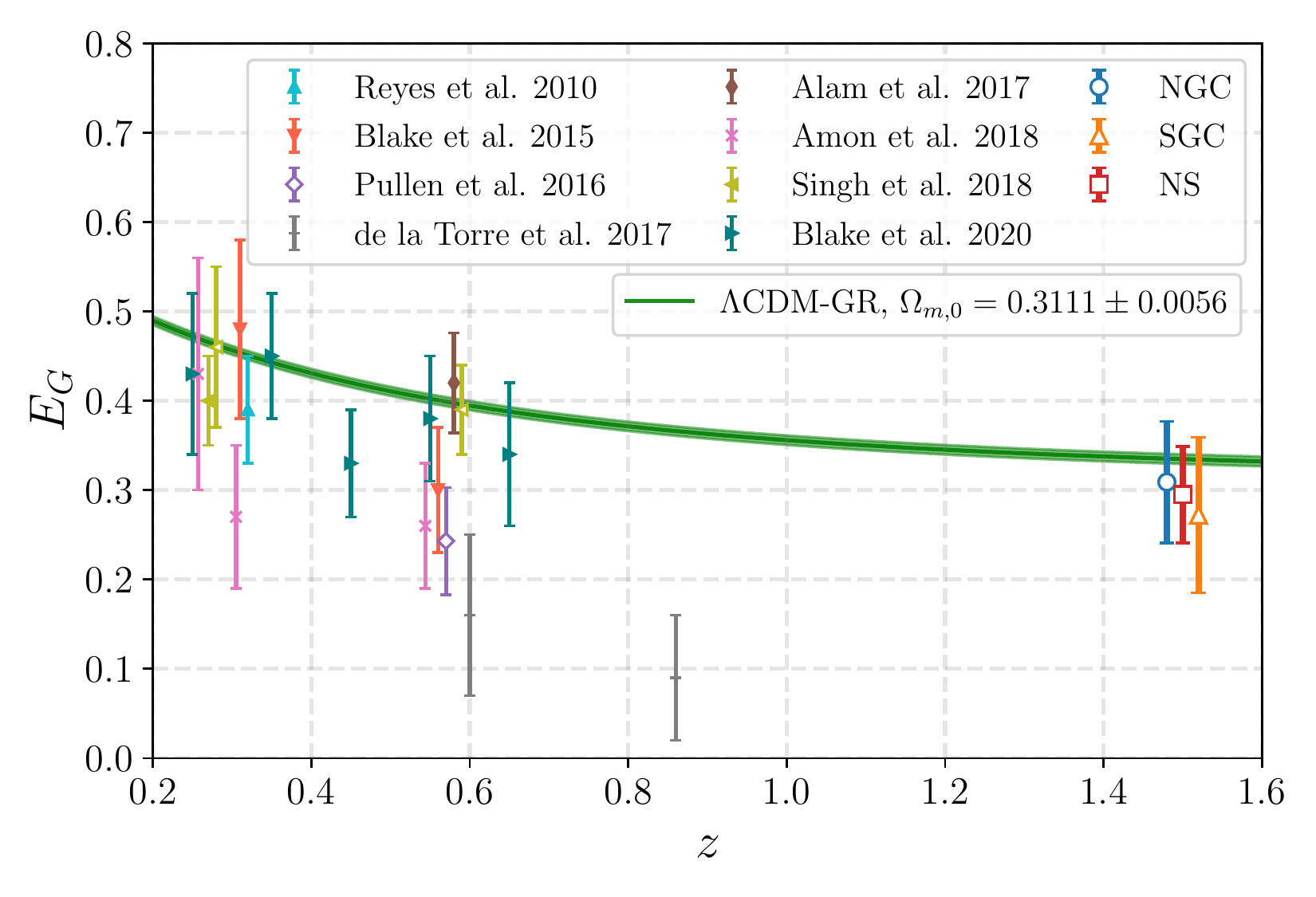}
    \caption{Some previous $E_G$ estimates and the results of this work. For reading convenience, some results are slightly shifted horizontally. For the results in this work, the NS is plotted at the effective redshift $z = 1.5$. The data points with white marker face color are estimated using CMB lensing while others are estimated with galaxy-galaxy lensing. The solid line is the $\Lambda$CDM-GR prediction (Eq.~\ref{eq:eg_gr}) with $\Omega_{\rm m,0}$ from \textit{Planck} 2018 CMB+BAO cosmological parameters.}
\label{fig:pre_egs}
\end{figure}
Some previous estimates of $E_G$ at lower redshifts and results in this work are summarized in Fig.~\ref{fig:pre_egs}. The statistical errors are still too large to discriminate between different gravity models. This work extends the redshift baseline of testing GR with $E_G$, while there is still a gap between $z\sim 0.6$ and $z\sim 1.5$, where $E_G$ has not been explored mainly due to the lack of promising LSS tracers considering the drop in the CMB lensing kernel.

There are still a few concerns which can be improved in the future with larger data samples. First, the redshift range $0.8 < z < 2.2$ of the quasar sample in this work is wide, and the effective redshift description may not be perfect. We tried to split the sample into smaller redshift bins, and study the redshift evolution of all the quantities. However, limited by the size of the sample, the SNRs are too low to give us reliable estimates. Second, we used both jackknife resampling and simulations to estimate the covariance matrix for $E_G(\ell)$, with the latter taken for the final result reported. However, we know that both these two methods have limitations. Although the simulations are designed to be realistic, it is still possible that there are unknown systematics that contribute to the covariances. For future surveys with a larger sky area, a larger number of jackknives would serve as a reliable comparison. At last, so far the statistical error bars are still very large, which make it difficult to do a selection of different gravity models. Also, a rigorous self-consistency test of any gravity model requires the corresponding fiducial cosmology and simulations. Besides, it is necessary to have simulated CMB lensing maps and galaxy/quasar mocks that are truly correlated for future surveys where lensing reconstruction and shot noise will be lower and the contribution to the covariance matrix from cross correlation will no longer be negligible.

\textit{Planck} has been a very successful CMB survey which gives the best constraints on the cosmological parameters so far. The next stage CMB surveys, e.g. CMB-S4~\citep{CMB-S4-2016} and Simons Observatory~\citep[SO;][]{SO-2019}, will produce even more accurate maps with higher resolution and lower noise. BOSS and eBOSS in SDSS has made the largest catalogs of LSS tracers in the Universe. While DR16 is the last data release of the series, more and larger LSS surveys are in progress. In the coming few years, the Dark Energy Spectroscopic Instrument~\citep[DESI;][]{DESI2016} survey will target about $17$ million ELGs in the redshift range $0.6\leq z\leq 1.6$, which will be able to fill the gap in Fig.~\ref{fig:pre_egs}. Redshifts of $1.7$ million quasars with $z<2.1$ as LSS tracers will also be measured over a sky area of $14\,000\,{\rm \deg}^2$, which corresponds to $f_{\rm sky}\simeq 34\,\%$. Compared with the eBOSS sample used in this work, the sky coverage and angular number density are increased by a factor of $3$ and $1.7$ respectively. Some analytic forecasts of constraining $E_G$ with future CMB and LSS surveys are discussed in \citet{Pullen2015}, where we can see that the SNR in this work can be improved by an order of magnitude with the DESI quasar sample. With all these promising future surveys, modern cosmology will be able to explore the origin and evolution of the Universe with higher and higher precision.

\section*{Acknowledgements}

We thank Jeremy Tinker and Michael Blanton for helpful discussions. We also thank eBOSS QGC working group for comments on the RSD fitting analysis. ARP was supported by NASA under award numbers 80NSSC18K1014 and NNH17ZDA001N. ARP was also supported by the Simons Foundation. SA is supported by the European Research Council through the COSFORM Research Grant (\#670193). ADM was supported by the U.S. Department of Energy, Office of Science, Office of High Energy Physics, under Award Number DE-SC0019022. GR acknowledges support from the National Research Foundation of Korea (NRF) through Grants No. 2017R1E1A1A01077508 and No. 2020R1A2C1005655 funded by the Korean Ministry of Education, Science and Technology (MoEST), and from the faculty research fund of Sejong University.

This work is based on observations obtained with \textit{Planck} (\url{http://www.esa.int/Planck}), an ESA science mission with instruments and contributions directly funded by ESA Member States, NASA, and Canada.

Funding for the Sloan Digital Sky Survey IV has been provided by the Alfred P. Sloan Foundation, the U.S. Department of Energy Office of Science, and the Participating Institutions. SDSS-IV acknowledges
support and resources from the Center for High-Performance Computing at
the University of Utah. The SDSS web site is www.sdss.org.

SDSS-IV is managed by the Astrophysical Research Consortium for the 
Participating Institutions of the SDSS Collaboration including the 
Brazilian Participation Group, the Carnegie Institution for Science, 
Carnegie Mellon University, the Chilean Participation Group, the French Participation Group, Harvard-Smithsonian Center for Astrophysics, 
Instituto de Astrof\'isica de Canarias, The Johns Hopkins University, Kavli Institute for the Physics and Mathematics of the Universe (IPMU) / 
University of Tokyo, the Korean Participation Group, Lawrence Berkeley National Laboratory, 
Leibniz Institut f\"ur Astrophysik Potsdam (AIP),  
Max-Planck-Institut f\"ur Astronomie (MPIA Heidelberg), 
Max-Planck-Institut f\"ur Astrophysik (MPA Garching), 
Max-Planck-Institut f\"ur Extraterrestrische Physik (MPE), 
National Astronomical Observatories of China, New Mexico State University, 
New York University, University of Notre Dame, 
Observat\'ario Nacional / MCTI, The Ohio State University, 
Pennsylvania State University, Shanghai Astronomical Observatory, 
United Kingdom Participation Group,
Universidad Nacional Aut\'onoma de M\'exico, University of Arizona, 
University of Colorado Boulder, University of Oxford, University of Portsmouth, 
University of Utah, University of Virginia, University of Washington, University of Wisconsin, 
Vanderbilt University, and Yale University.

\section*{Data availability}
The eBOSS DR16 quasar clustering catalogs are available in the SDSS-IV Science Archive Server at \url{https://data.sdss.org/sas/dr16/eboss/lss/catalogs/DR16/}.
The \textit{Planck} 2018 CMB lensing data and simulations~\footnote{\url{https://wiki.cosmos.esa.int/planck-legacy-archive/index.php/Lensing}} were accessed from Planck Legacy Archive at \url{https://pla.esac.esa.int/}.
The derived data generated in this research will be shared on reasonable request to the corresponding author.




\bibliographystyle{mnras}
\bibliography{eg_qsoxcmb}




\appendix

\section{CMB contamination systematic bias} \label{sec:cmb-sys}
\begin{figure*}
    \centering
    \includegraphics[width=\textwidth]{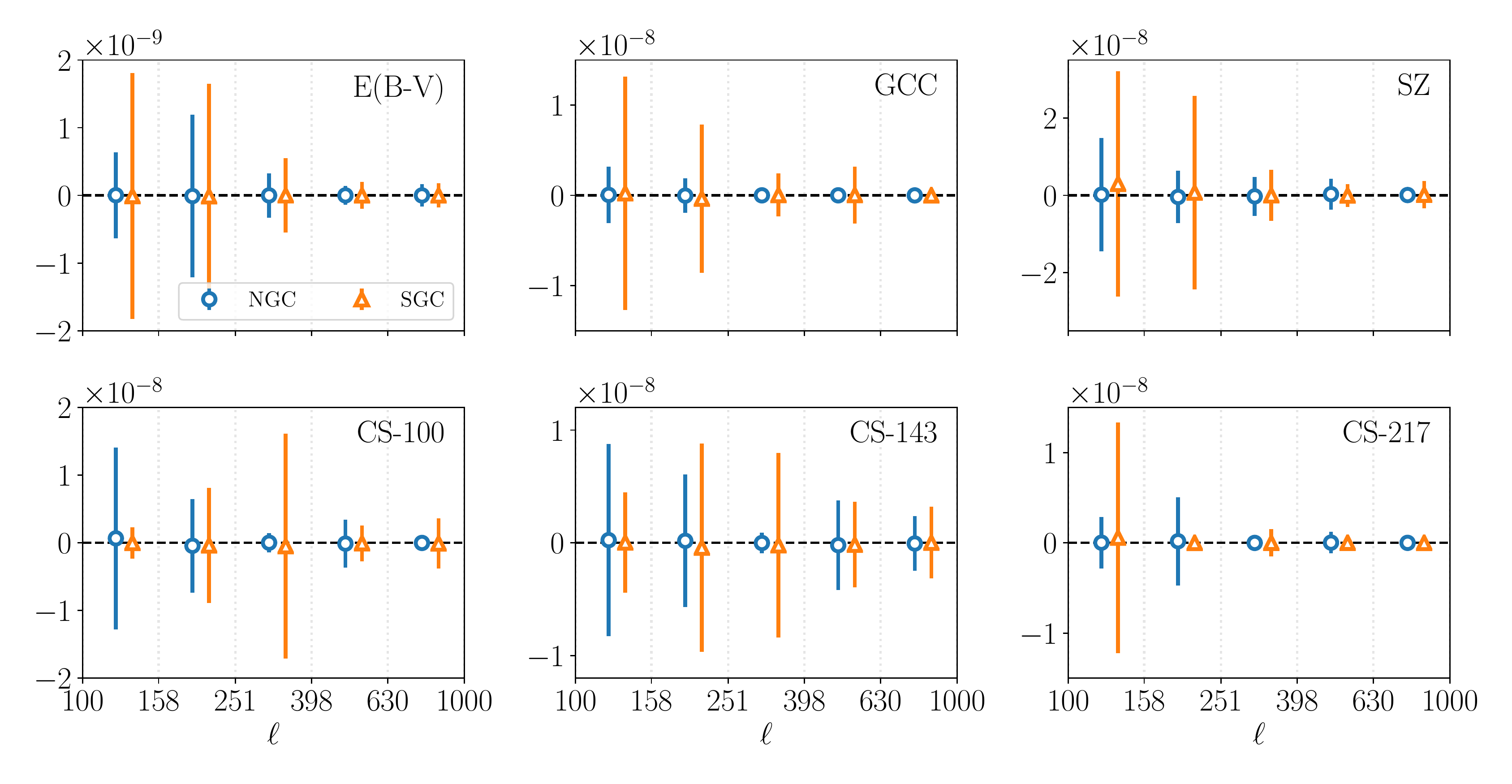}
    \caption{The estimated bias $\Delta\hat{C}_\ell^{\kappa q}$ to the CMB lensing $\times$ quasar cross-power spectrum caused by possible contamination sources in the CMB temperature map.}
    \label{fig:sys}
\end{figure*}
In this section, we consider the foregrounds, including dust and point sources, that might contaminate the CMB temperature maps. These sources can leave signatures in the CMB lensing map and hence bias our cross-correlation signal with quasars.

We use the galactic dust emission map constructed in~\citet{Schlegel1998}. For point sources, we make angular maps for several \textit{Planck} point source catalogs, including galactic cold clumps~\citep[GCC;][]{Planck2015-CGCC}, Sunyaev-Zel'dovich~\citep[SZ;][]{Sunyaev1980} sources~\citep{Planck2015-2ndCSZs} and compact sources~\citep[CS;][]{Planck2015-2ndCCS} at $100$, $143$ and $217$ GHz. Following~\citet{Pullen2016}, we estimate the biases to $\hat{C}_\ell^{\kappa q}$ by these possible sources with
\begin{equation}
    \Delta \hat{C}_{\ell}^{\kappa q} = \frac{\hat{C}_\ell^{\kappa c}\hat{C}_\ell^{qc}}{\hat{C}_\ell^{cc}}\, ,
\end{equation}
where $c$ is any of the contamination maps, and the errors are given by
\begin{equation}
    \sigma^2\left({\Delta \hat{C}_{\ell}^{\kappa q}}\right) = \left(\Delta \hat{C}_{\ell}^{\kappa q}\right)^2 \left[\frac{\sigma^2(\hat{C}_\ell^{\kappa c})}{(\hat{C}_\ell^{\kappa c})^2} + \frac{\sigma^2(\hat{C}_\ell^{q c})}{(\hat{C}_\ell^{q c})^2}\right]\, .
\end{equation}
The estimates are shown in Fig.~\ref{fig:sys}.
We find that the biases are consistent with zero, with statistical errors that are much lower than our $\hat{C}_\ell^{\kappa q}$ signal (Fig.~\ref{fig:clkq}). This is expected since the most foreground-contaminated area of the dust emission map, i.e. the Galactic plane, and the sky regions of many point sources have already been masked out in the \textit{Planck} maps. Compared with the previous analysis for cross-correlating CMASS galaxies~\citep{Pullen2016} with \textit{Planck} 2015 CMB lensing map, the removal of the contamination has been improved for \textit{Planck} 2018 data release. A similar analysis has also been conducted for eBOSS DR14 quasars~\citep{Han2019}.

\section{Test on fitting \texorpdfstring{$E_G(\ell)$}{EG(ell)} over scales} \label{sec:rl}
Here we take a slightly different approach on fitting $E_G(\ell)$ over scales (i.e. the $5$ bandpowers) for the scale-averaged $\bar{E}_G$, which also serves as a test on the reliability of estimating the covariance matrices (CMs) with simulations and jackknife resampling.

\begin{figure}
    \centering
    \includegraphics[width=\columnwidth]{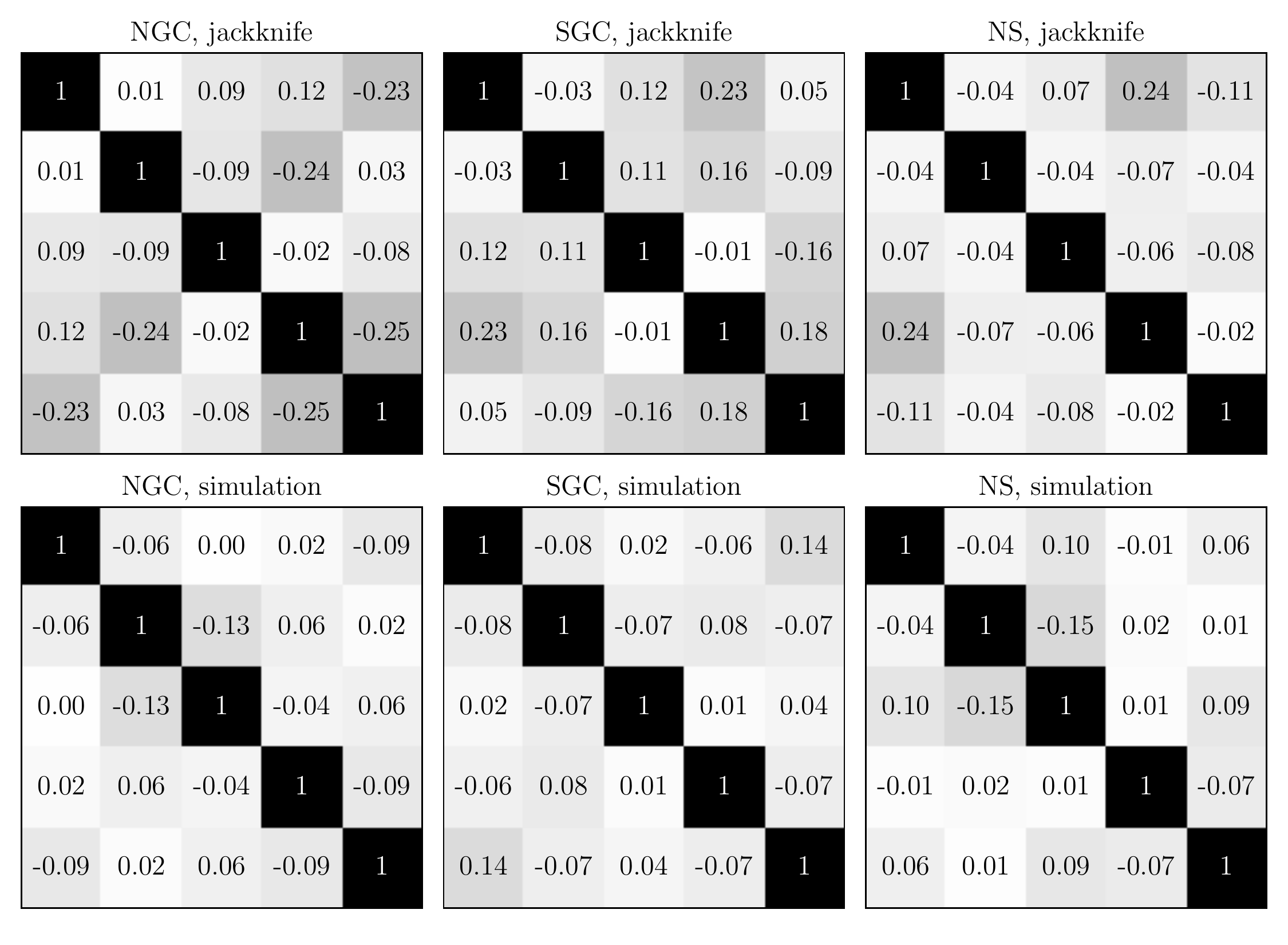}
    \caption{Estimated correlation matrices, similar as Fig.~\ref{fig:eg_cov} but for $R_\ell$ (Eq.~\ref{eq:rl}).}
    \label{fig:rl_cov}
\end{figure}

With our $E_G(\ell)$ estimator given by Eq.~\ref{eq:eg_est} and assuming a scale-independent RSD parameter $\beta$, $E_G(\ell)$ could be scale-dependent only through the ratio of the angular power spectra,
\begin{equation}
    R_\ell \equiv C_\ell^{\kappa q} / C_\ell^{qq} \,.
\label{eq:rl}
\end{equation}
Thus fitting $E_G(\ell)$ as discussed in Section~\ref{subsec:eg_res} should be equivalent to fitting $R_\ell$ over scales first, whose best-fit estimate is then combined with $\beta$ into $\bar{E}_G$. The Gaussian likelihood function and best-fit value are in the same form as that for $E_G(\ell)$ (Eq.~\ref{eq:eg_likelihood} and~\ref{eq:eg_bestfit}), with $E_G(\ell)$ replaced by $R_\ell$. The key point is that the corresponding $\hat{\mathbf{C}}$ is now the CM for $R_\ell$.
\begin{figure}
    \centering
    \includegraphics[width=\columnwidth]{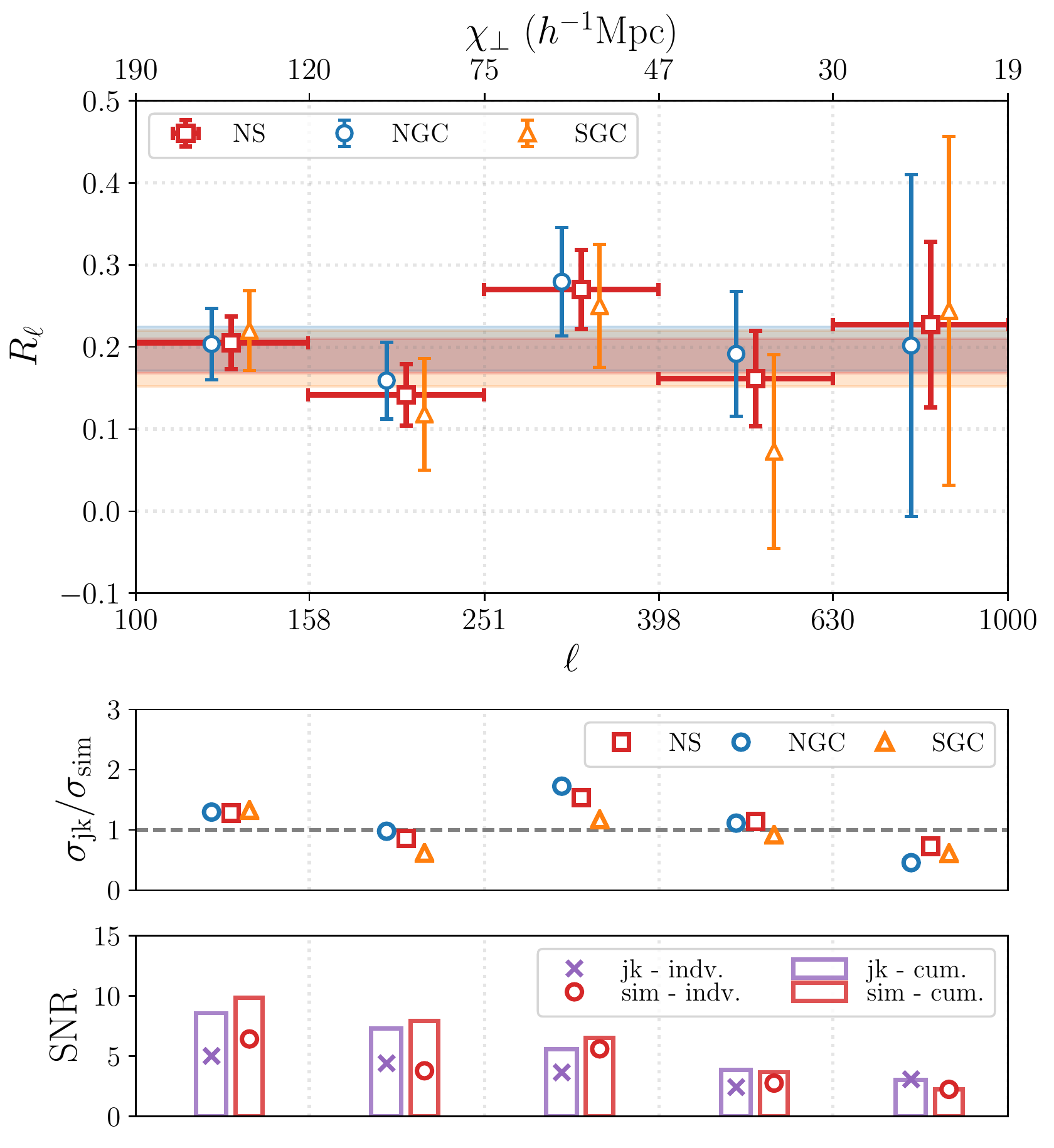}
    \caption{$R_\ell$ (Eq.~\ref{eq:rl}) estimates, with similar information as in Fig.~\ref{fig:eg}. The shaded area in the \textit{upper} panel is the best-fit value over scales with $1\,\sigma$ error, for each of the two caps and the combination.}
    \label{fig:rl}
\end{figure}
We present the estimates and the correlation matrices of $R_\ell$ in Fig.~\ref{fig:rl} and~\ref{fig:rl_cov}.
Compared with that for $E_G(\ell)$ (Fig.~\ref{fig:eg_cov}), the cross correlations between scales are weaker, which is expected since using the same scale-independent $\beta$ value for all bins of $E_G(\ell)$ introduces covariances.
\begin{table}
    \centering
    \caption{Scale-averaged $E_G$ estimates, similar as Table~\ref{tab:eg_ests} but with the approach discussed in Appendix~\ref{sec:rl}.}
    \label{tab:eg_ests_rl}
    \begin{tabular}{cccc}
        \hline
        Cap & NS & NGC & SGC \\
        $E_G$ with $\hat{\mathbf{C}}_{\rm sim}$ & $0.294\pm 0.057$ & $0.308\pm 0.073$ & $0.272\pm 0.092$ \\
        $E_G$ with $\hat{\mathbf{C}}_{\rm jk}$ & $0.267\pm 0.045$ & $0.291\pm 0.062$ & $0.240\pm 0.066$ \\
        \hline
    \end{tabular}
\end{table}
With the $C_\Gamma$ (Eq.~\ref{eq:gamma_cor}, which are almost the same value for the $5$ bins) calibration factor applied, we summarize the final scale-independent $\bar{E}_G$ estimates with this second approach in Table~\ref{tab:eg_ests_rl}. Compared with the results in Table~\ref{tab:eg_ests}, the estimates with simulated CMs are well consistent while those with jackknife resampling CMs differ by $3\%$ for NGC and $12\%$ for SGC. This disagreement in jackknife resampling can be caused by the small number of samples, which may not be enough to give us accurate CMs for either or both of $E_G(\ell)$ and $R_\ell$. On the other hand, for simulations, the consistency between the two approaches indicates that the CMs should be well constrained.


\bsp	
\label{lastpage}
\end{document}